\documentclass[prb,aps,twocolumn,groupaddress,nofootinbib,floatfix]{revtex4-1}
\usepackage{latexsym}
\usepackage{graphicx}
\usepackage{verbatim}
\usepackage{multirow}
\usepackage{amsmath}
\usepackage{mathrsfs}
\usepackage{float}
\renewcommand{\vec}[1]{{\mathbf{#1}}}
\usepackage[usenames, dvipsnames]{color}
\usepackage{mathtools}
\usepackage{pgfplots}
\pgfplotsset{compat=1.4}
\usepgfplotslibrary{groupplots}

\usepackage{slashed}
\usepackage{physics}	
\usepackage{graphicx} 
\usepackage{epstopdf}
\usepackage{bbold}
\usepackage{wasysym}

\usepackage[ampersand]{easylist}

\newcommand{\beq}{\begin{eqnarray}}
\newcommand{\eeq}{\end{eqnarray}}

\newcommand \adj {^\dagger}
\newcommand \inv {^{-1}}
\renewcommand \hat [1] {\widehat{#1}}
\renewcommand \bar [1] {\overline{#1}}

\newcommand \NN {\nonumber}
\renewcommand \P [1]{\left(#1\right)}
\newcommand \B [1]{\left[#1\right]}

\newcommand \ave [1]{\langle{#1}\rangle}

\newcommand \x {\times}

\renewcommand \tr {\operatorname*{tr}}

\newcommand \pd[3][] {\frac{\partial^{#1}#2}{\partial{#3}^{#1}}}

\newcommand \ph {{\mathrm{ph}}}

\newcommand \HK {{\mathrm{HK}}}

\newcommand \up {\uparrow}
\newcommand \dn {\downarrow}

\usepackage[final,unicode,colorlinks=true,citecolor=purple,linkcolor=blue,linktocpage]{hyperref}
\usepackage[obeyFinal,textsize=tiny,backgroundcolor=white,linecolor=magenta]{todonotes}

\begin{document}

\title{Thermodynamics of an Exactly Solvable Model for Superconductivity in a Doped Mott Insulator}

\author{Jinchao Zhao}
\author{Luke Yeo}
\author{Edwin W. Huang}
\author{Philip W. Phillips}
\affiliation{Department of Physics and Institute of Condensed Matter Theory, University of Illinois at Urbana-Champaign, Urbana, 61801 Illinois, USA}

\begin{abstract}
Computing superconducting properties starting from an exactly solvable model for a doped Mott insulator stands as a grand challenge. We have recently shown that this can be done starting from the Hatsugai-Kohmoto (HK) model which can be understood generally as the minimal model that breaks the non-local $\mathbb Z_2$ symmetry of a Fermi liquid, thereby constituting a new quartic fixed point for Mott physics [Phillips et al., Nature Physics 16, 1175 (2020); Huang et al., Nature Physics (2022)]. In the current work, we compute the thermodynamics, condensation energy, and electronic properties such as the NMR relaxation rate $1/T_1$ and ultrasonic attenuation rate. Key differences arise  with the standard BCS analysis from a Fermi liquid: 1) the free energy exhibits a local minimum at $T_p$ where the pairing gap turns on discontinuously above a critical value of the repulsive HK interaction, thereby indicating a first-order transition,  2) a tri-critical point emerges, thereby demarcating the boundary between the standard second-order superconducting transition and the novel first-order regime,  3) Mottness changes the sign of the quartic coefficient in the Landau-Ginzburg free-energy fuctional relative to that in BCS, 4) as this obtains in the strongly interacting regime, it is Mott physics that underlies the generic first-order transition, 5) the condensation energy exceeds that in BCS theory suggesting that multiple Mott bands might be a way of enhancing superconducting, 6) the heat-capacity jump is non-universal and increases with the Mott scale, 7) Mottness destroys the Hebel-Slichter peak in NMR, and 8) Mottness enhances the fall-off of the ultrasonic attenuation at the pairing temperature $T_p$.  As several of these properties are observed in the cuprates, our analysis here points a way forward in computing superconducting properties of strongly correlated electron matter.  
\end{abstract}

\maketitle

\section{Introduction}

A truly remarkable feature of superconductivity in elemental metals is that although the superconducting order parameter, $\Psi=\Delta e^{i\theta}$, has two components, a phase, $\theta$ and an amplitude, $\Delta$, both turn on at the same temperature. This behaviour is captured by the mean-field treatment of BCS\cite{BCS} which predicts a second-order phase transition in which superconducting fluctuations of the pair amplitude satisfy the Ginzburg criterion, $\delta (\Delta)^2/\Delta\ll 1$. Should this criterion be met, pairing and phase coherence are synonymous. From a technical standpoint, satisfying the Ginzburg criterion is quite surprising as the latter is governed by a divergence in the pair susceptibility while the former stems from a solution to a mean-field integral equation. Experimentally, no more than a 1K difference between these temperatures in BCS materials is observed. Hence, the Ginzburg criterion is really a statement about the accuracy of mean-field theory. 

It is well known that the cuprates violate the Ginzburg criterion or equivalently the BCS dictum that pairing without long-range order is impossible\cite{bozovic1,bozovic2,uemura,kivelem}. While this trend was thought to vanish in the overdoped regime, recent spectroscopic measurements\cite{birgeneau} on the bilayer cuprate (Pb,Bi)$_2$Sr$_2$CaCu$_2$O$_{8+\delta}$ (Bi-2212) have verified that even in such samples remnants of a particle-hole symmetric superconducting gap persist up to $T_{\rm pair}=86K$ where the superconducting transition is $T_c=63K$. The focus on a particle-hole symmetric gap as the signature of the pairing gap is designed to disentangle superconductivity-related gaps from the wider range of phenomena associated with the pseudogap\cite{norman} that need not\cite{lanzarapha,pickettpha,AHO4} have this symmetry. Within the cuprate family, Bi-2212 and YBa$_2$Cu$_3$O$_{7-\delta}$ (YBCO) have the highest ratio of $T_{\rm pair}/T_c$ of 1.6 while in (La,Sr)$_2$CuO$_4$ (LSCO) the ratio is 1.2. Although 2D disordered materials are expected to have a wide range where thermal superconducting fluctuations obtain\cite{goldman,yazdani}, the efficient cause of the disconnect between phase coherence and the gap turn-on remains unsettled in the cuprates. For example, the discrepancy has been attributed\cite{birgeneau} to the existence of a flat band at $(\pi,0)$. Further, it is unclear how to think about a pairing gap without a phase transition. The question arises, what is the order of the transition in which pairing and phase coherence are decoupled? It is questions of this type that we address in this work.

In trying to understand the source of the discrepancy between pairing and phase coherence, it is worth cataloguing other instances of deviations from BCS superconductivity in the cuprates. Two features stand out: 1) the color change\cite{marelcolor} and 2) a violation of the Glover-Ferrell-Tinkham\cite{FGT1,FGT2} (GFT) sum rule. Regarding the latter, in a standard BCS superconductor, condensation leads to loss of spectral weight
at energy scales no more than ten times the pairing energy. Not so in the cuprates. Bontemps and colleagues\cite{bontemps} have
directly observed that in underdoped (but not overdoped) Bi-2212, the Glover-Ferrel-Tinkham sum rule is violated
and the optical conductivity must be integrated to 20,000$cm^{-1}$
to recover the spectral weight lost upon condensation
into the superconducting state. Similarly, R\"ubhaussen, et. al\cite{rubhaussen} and others\cite{homes} have shown that changes in the optical conductivity occur at energies 3eV (roughly
100$\Delta$ where $\Delta$ is the maximum superconducting gap) away from the Fermi energy at $T_c$. Finally, van der Marel and colleagues\cite{marelcolor} have seen 
an acceleration in the depletion of the high energy spectral weight accompanied with a compensating increase in
the low-energy spectral weight at and below the superconducting transition. Specifically the integrated weight of the optical conductivity over the lower Hubbard band scales ($<1eV$) increases below the superconducting temperature, whereas the high-energy component ($[1eV,2eV]$) decreases. Since the integrated weight determines the number of charge carriers, the color change indicates that high-energy scales contribute to the superfluid density in contrast to the standard BCS picture. Consequently, the superfluid density in the cuprates is derived not just from low energy physics. In essence, it involves UV-IR mixing.

What all of this seems to indicate is that the departures from BCS superconductivity in the cuprates are tethered to the Mott state. What is difficult then is to solve a reasonable model for a Mott insulator which captures this range of deviations from the BCS paradigm. It is in attempting to answer this array of questions that we have focused\cite{hksupercon} on the Hatsugai-Kohmoto (HK)\cite{HKmodel} model, an exactly solvable model for a Mott insulator. The HK model,
\beq
\label{HKham}
H_{HK}=\sum_{\vec k\sigma}\xi_{\vec k} n_{\vec k\sigma}+U\sum_{\vec k} n_{\vec k\uparrow}n_{\vec k\downarrow},
\eeq
is essentially the Hubbard model in momentum space. As we pointed out previously\cite{HKrenorm}, this model is important because it represents the simplest way of breaking the hidden $\mathbb Z_2$ symmetry of a Fermi liquid. We illustrate this from the basic Hamiltonian for a Fermi liquid
\beq
\label{free}
H_{\rm FL}=\sum_{\vec p} \psi_{\vec p}^\dagger (\epsilon_{\vec p}-\epsilon_F)\tau_3\psi_{\vec p}+\cdots.
\eeq
Here $\psi^\dagger_{\vec p}=(c^\dagger_{\vec p\uparrow},c_{-\vec p\downarrow})$ and $\tau_3$ is the standard z-Pauli matrix.
For electrons at the Fermi surface, $\epsilon_{\vec p}=\epsilon_F$, this Hamiltonian obeys the symmetry $n_{\vec k\uparrow}\rightarrow-n_{\vec k\uparrow}$ where only one of the spin currents changes sign. Consequently, the interaction term of the form in the HK model maximally breaks this symmetry. Based on this, we showed previously\cite{HKrenorm} that the HK model represents a fixed point for Mott physics that even encompasses Hubbardology. A straightforward Fourier transform of the Hubbard on-site interaction reveals that it contains the HK interaction term. As we have shown previously\cite{HKrenorm}, it is this term that is most relevant in the renormalization sense and the only one that maximally breaks the hidden $Z_2$ symmetry of a Fermi liquid. For these reasons, we have focused on revealing its superconducting properties\cite{hksupercon}.  To accomplish this, we appended to the HK Hamiltonian, which only describes a doped Mott insulator, a pairing term, in analogy with the Cooper program\cite{BCS} in which a pairing term was added to a Fermi liquid, to investigate if a corresponding pairing instability obtains. Indeed it did\cite{hksupercon} and hence we found a computationally tractable starting point for superconductivity and Mottness.  While our previous work revealed $T=0$ properties of the HK model appended with a pairing term with couplinng constant, $g$, none of the thermodynamics were obtained. Nonetheless, several non-BCS properties were apparent: 1) $\lim_{g\rightarrow 0} \frac{2\Delta}{T_c}\rightarrow \infty$ rather than the BCS ratio of $3.52$, 2) Composite quasiparticle excitations consisting of doublons and holons rather than the standard particle-hole excitations of BCS, and 3) a suppression of the superfluid density relative to that of BCS. 

Because the model we explore is highly amenable to computation, we can with certainty catalog how the finite temperature properties derived from Mottness lead to substantial deviations from BCS theory. First, we establish that the turn-on of the gap and the divergence of the susceptibility occur at distinct temperatures. The former appears to obtain at a first-order transition while the latter tends to a global second-order transition of the superconducting state. We show that this effect vanishes when $U=0$, thereby making this a true consequence of Mottness. We make contact with earlier results on multi-band superconductors that found a first-order transition\cite{FOT1,FOT2,FOT3,FOT4}. We trace the first-order nature of the pairing transition to a singly-occupied holon band that arises purely from Mott physics. Second, we compute the heat capacity jump at $T_c$ as well as the condensation energy. Unlike BCS theory, we find that $E_{\rm cond}/\Delta$ is stronger than in BCS theory. Finally, we compute the ultrasonic attenuation as well as $T_1$ near the superconducting transition. We are able to show that the Hebel-Slichter\cite{HS} peak in BCS theory vanishes in the strongly correlated limit as seen widely in the cuprates\cite{NMR1,NMR2,NMR3}. Although the absence of this peak in the cuprates\cite{HSpeak,HSpeak2} has been attributed to spin fluctuations, we argue here that it is just a consequence of Mottness, the splitting of the spectral weight over two correlated bands. Subsequent experiments are discussed.  While it is possible to use this model to address a possible BCS/BEC crossover, we do not explore this here.  Such a crossover has been explored previously from a Luttinger surface\cite{BEC/BCS}.

\section{Superconductivity in the HK model}

Superconductivity in the cuprates necessitates a solution to at least the Cooper instability in a doped Mott insulator. We have shown\cite{hksupercon} previously that this can be done exactly by solving Cooper instability equation that arises from the HK analogue 
\beq
H = H_\HK - H_p,
\qquad H_p = \frac{g}{V} \sum_{\vec k,\vec k'}b_{\vec k}\adj b_{\vec k'}
\label{Htot}
\eeq
of the pairing Hamiltonian for a doped Mott insulator. Here $b_{\vec k}= c_{-\vec k\downarrow} c_{\vec k\uparrow}$ is the $s$-wave pair creation operator at zero total momentum. 
As is well known for Mott systems, the single-particle Green function for $H_{\rm HK}$,
\begin{align}\label{eq:propagator}
 G_{\vec k\sigma}(i\omega_n)
 &\equiv -\int_{0}^{\beta} d\tau\; \ave{c_{\vec k\sigma}(\tau) c_{\vec k\sigma}\adj(0)} e^{i\omega_n\tau} \\
 G_{\vec k\sigma}(i\omega_n \to z) &= \frac{1-\ave{n_{\vec k\bar{\sigma}}}}{z - \xi_{\vec k}} + \frac{\ave{n_{\vec k\bar{\sigma}}}}{z - (\xi_{\vec k} + U)}. 
\end{align} 
exhibits a bifurcation of the spectral weight between lower (l) and upper (u) bands with weights that are determined by the electron filling, $1-\langle n_{\vec k\bar{\sigma}}\rangle$ and $\langle n_{\vec k\bar{\sigma}}\rangle$, respectively. It is this bifurcation that leads to zeros of the real part of the Green function.  The features in the lower band are created with the momentum-projected operators $\xi_{k\sigma}=c_{k\sigma}(1-n_{k\bar{\sigma}})$.  In the Hubbard model, the corresponding operators for the lower band are not known exactly.  With these operators, it is easy to see that the physics of the lower Hubbard band in the HK model is not that of a Fermi liquid.  Namely,  there are excited states of the HK model that have no correspondence with those in a FL.  Consider a two-particle excitation.  This would be generated by applying the $\xi^\dagger_{k\sigma} $ operator twice.  That is, the excited states  should be described by $\xi^\dagger_{k\uparrow}\xi^\dagger_{k\downarrow}$.  However, this operator is explicitly zero.   That is, there are excitations in the HK model that have no counterpart by acting with the single-particle operators.  At work here is the fact that as long as $U\ne 0$, a Luttinger surface of zeros is present.  There are strictly no pure pole-like excitations.  Hence, there is no FL part of the HK model as long as $U\ne 0$. 

We have also shown\cite{hksupercon} that not only is the Cooper instability exactly solvable but so is the exact pair susceptibility. The exact susceptibility\cite{hksupercon} 
\beq
 \chi(i\nu_n) \equiv \frac{1}{V} \sum_{\vec k,\vec k'}\int_{0}^{\beta} d\tau\; e^{i\nu_n \tau} \ave{T b_{\vec k}(\tau) b_{\vec k'}\adj}_{g}
 \eeq
 can be expressed, in the normal state, in terms of the bare susceptibility 
 \beq
 \chi(i\nu_n)
 = \frac{\chi_0(i\nu_n)}{1 - g \chi_0(i\nu_n)}
\eeq
which is given by
\begin{align}
 \chi_0(i\nu_n) &= \chi_0^{l l} + \chi_0^{u u} + \chi_0^{l u} + \chi_0^{u l} \\
 \chi_0^{a b} &= \frac{1}{V} \sum_{\vec k} n_{\vec k\up}^a n_{-\vec k\dn}^b \frac{f(\omega_{\vec k}^a)+f(\omega_{-\vec k}^b)-1}{i\nu_n-\omega_{\vec k}^a-\omega_{-\vec k}^b}
\end{align}
where for $\omega_{\vec k}^l = \xi_{\vec k}$ and $\omega_{\vec k}^u = \xi_{\vec k}+U$, $n_{\vec k\sigma}^u = \ave{n_{\vec k\bar{\sigma}}}_0$ and $n_{\vec k\sigma}^l = 1-n_{\vec k\sigma}^u$, and $f(\omega)$ the Fermi function at temperature $T$, the superscripts $a b$ may represent $l l$, $u u$, $l u$, or $u l$.

A consequence of the expression of the pair susceptibility is that the divergence at $\chi_0=1/g$ is expected to be a second-order transition to the bulk superconducting state. As we will see, this divergence is not coincident with the turn-on of the gap. To calculate the susceptiblity, we will work with the exact finite temperature occupancy, 
\beq
\ave{n_{\vec k\sigma}}=\frac{1}{2}\ave{n_{\vec k}}=\frac{e^{-\beta\xi_{\vec k}}+e^{-\beta(2\xi_{\vec k}+U)}}{1+2e^{-\beta\xi_{\vec k}}+e^{-\beta(2\xi_{\vec k}+U)}},
\eeq
so as to give the correct temperature dependence of $\chi(T)$ explicitly. 
\begin{figure}[ht]
 \centering
 \includegraphics[width=8.6cm]{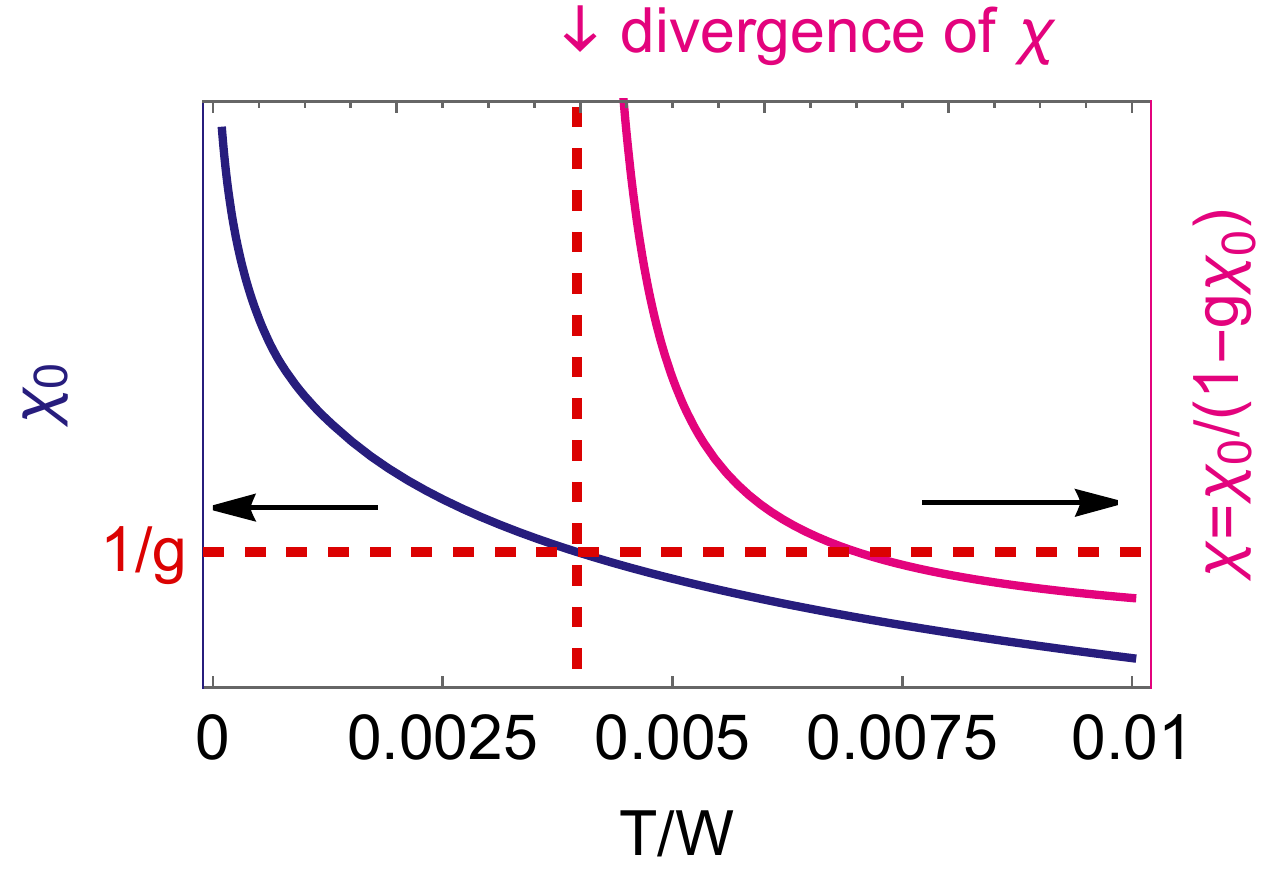}
 \caption{The temperature dependence of $\chi_0$(left) and $\chi$(right) for a 3D-HK model with superconducting pairing. The energy dispersion was approximated by a parabola dispersion, and the chemical potential was set to half filling of the lower Hubbard band, $\mu=0.5W$ where $W$ is the bandwidth of the lower Hubbard band. We take $W=1$ in the following calculation. The temperature $T_c$ at which $\chi=\chi_0/(1-g\chi_0)$ diverges ($\chi_0=1/g$) represents the true superconducting transition onset.}
 \label{fig:susceptibility}
\end{figure}
In Fig. \ref{fig:susceptibility}, we plot the zero-frequency bare susceptibility $\chi_0(T)$ as well as $\chi(T)$. The divergence of $\chi(T=T_c)$ would imply a diverging length scale and thus a possible second-order phase transition temperature $T_c$. Above $T_c$ there is no extra sigularity in the susceptibility, which is the same case as in the BCS theory. To reiterate, the onset of the gap, a mean-field notion, is distinct from the divergence of the susceptibility.

\subsection{Mean-field Theory of HK}

Given that the susceptibility calculation is exact, we can assert that the ground state of the HK model with a pairing term is superconducting. To describe this state, we resort to a mean-field description in the spirit of BCS. In our previous work, we implemented the mean-field by an appropriately chosen pairing wave function. To go beyond such a ground-state treatment, we explicitly diagonalize $H$, obtaining all eigenstates and measuring observables in grand canonical ensemble, including their full temperature dependence.
The procedure we use has been outlined in the Supplementary Materials of Ref.~\onlinecite{hksupercon} and by Zhu, et al.\cite{Zhu2021}. In constructing the mean-field, we take advantage that the HK Hamiltonian does not mix the various k-states. In terms of the pair amplitude, $\Delta\equiv (g/V)\sum_{\vec k}b_{\vec k}$, we formulate the mean-field,
\begin{equation}
\begin{split}
 H_p&=\frac{g}{V}\sum_{\vec k\vec k'}b_{\vec k}^\dagger b_{\vec k'}\\
 &=\frac{V}{g}\Delta^\dagger\Delta\\
 &=\frac{V}{g}(\delta\Delta^\dagger+\bar\Delta^*)(\delta\Delta+\bar\Delta)\\
 &\approx\sum_{\vec k}(\bar\Delta b_{\vec k}^\dagger+\bar\Delta^*b_{\vec k})-\frac{V}{g}|\bar\Delta|^2,
\end{split}
\end{equation}
entirely on the pairing term where we have introduced the average of the pairing amplitude to be $\bar\Delta=\langle\Delta\rangle$.
In the last step, mean-field amounts to ignoring the second-order fluctuation $O(\delta\Delta^2)$ term. 
The mixing between different momentum sectors averages out because of the momentum-diagonal structure of HK. Thus the mean-field(MF) HK Hamiltonian can be block diagonalized\cite{hksupercon,Zhu2021}
\begin{equation}
\begin{split}
 H&\approx \sum_{\vec k\in\rm{HFBZ}}H^{MF}_{\vec k}+\frac{V}{g}|\bar\Delta|^2\\
 H^{MF}_{\vec k}&=\sum_{s=\pm}\xi_{\vec k}\sum_\sigma c_{s\vec k\sigma}^\dagger c_{s\vec k\sigma}+Uc_{s\vec k\uparrow}^\dagger c_{s\vec k\uparrow}c_{s\vec k\downarrow}^\dagger c_{s\vec k\downarrow}\\
 &\qquad-\left(\bar\Delta^*c_{-s\vec k\downarrow}c_{s\vec k\uparrow}+\bar\Delta c_{s\vec k\uparrow}^\dagger c_{-s\vec k\downarrow}^\dagger\right).
\end{split}
\end{equation}
The summation over k is carried out inside half of the first Brillouin zone(HFBZ), while the boundary terms with momenta restricted to the edge of the first Brillouin zone are neglected as they are suppressed by a factor of $1/N$ and hence vanish in the thermodynamic limit. 

The decomposed Hamiltonian lives in a Fock space $F_{\vec k}$ which contains 4 fermion currents, $\ket{n_{\vec k\uparrow},n_{\vec k\downarrow},n_{-\vec k\uparrow},n_{-\vec k\downarrow}}$ and spans a 16-dimensional space\cite{Zhu2021}. Due to the fermion parity conservation of $H_{\vec k}^{MF}$, this Fock space can be decomposed by parity into $F_{\vec k}=F_{\vec k}^{odd}\oplus F_{\vec k}^{even}$. The even sector is further block diagonalized into 3 subspaces, $F_{\vec k}^{even}=F_{\vec k}^{PB}\oplus F_{\vec k}^{ST}\oplus F_{\vec k}^{Mix}$, as shown in Table \ref{tab:energylevels} :
\begin{table}[t]
 \centering
 \begin{tabular}{c|c|c|c}
 \hline
  subspace & eigenvalue&degeneracy & basis \\
  \hline
  PB& $E_{\vec k}^4\equiv2\xi_{\vec k}+U$&2 & $\ket{1100},\ket{0011}$\\
  \hline
  ST& $E_{\vec k}^5\equiv 2\xi_{\vec k}$&3 & 
  $\begin{array}{c}
  \ket{1010},
  \ket{0101},\\
  (\ket{1001}-\ket{0110})/\sqrt2\\
  \end{array}$\\
  \hline
  Mixing& 
  $\begin{array}{c}
  E_{\vec k}^1\equiv E_1 \\
  E_{\vec k}^2\equiv E_2 \\
  E_{\vec k}^3\equiv E_3\\
  \end{array}$&
  $\begin{array}{c}
  1 \\
  1 \\
  1
  \end{array}$
  &
  $\begin{array}{c}
  \ket{0000}, \\
  (\ket{1001}+\ket{0110})/\sqrt2,\\
  \ket{1111} 
  \end{array}$\\
  \hline
  Odd$\times4$&
  $\begin{array}{c}
  E_{\vec k}^6\equiv E_- \\
  E_{\vec k}^7\equiv E_+ 
  \end{array}$&
  $\begin{array}{c}
  4 \\
  4 
  \end{array}$
  &
  $\begin{array}{c}
  \{\ket{1000},\ket{1110}\} \\
  \{\ket{0100},\ket{1101}\}\\
  \{\ket{0010},\ket{1011}\}\\
  \{\ket{0001},\ket{1011}\}\\
  \end{array}$\\
  \hline
 \end{tabular}
 \caption{Block decomposition and the energy levels of the HKSC mean-field hamiltonian}
 \label{tab:energylevels}
\end{table}
\begin{enumerate}
 \item The states in the Pauli Blocking (PB) states, possess an energy level at $E_{\vec k}^4=2\xi_{\vec k}+U$ with degeneracy 2, which have definite total electron number 2(not participating the superconducting pairing) and is blocked by the repulsion $U$. 
 \item The states in the Spin Triplet (ST) states, possess an energy level at $E_{\vec k}^5=2\xi_{\vec k}$ with degeneracy 3, which also have two electrons(not participating the superconducting pairing).
 \item The 3-dimensional particle number mixing (Mixing) states related by the off-diagonal superconducting pairing.
\end{enumerate}
The Hamiltonian matrix in the 3-dimensional mixing states subspace is
\beq
 \left(
 \begin{array}{ccc}
  0&-\sqrt{2\bar\Delta} &0 \\
  -\sqrt{2\bar\Delta}&2\xi_{\vec k}&-\sqrt{2\bar\Delta}\\
  0&-\sqrt{2\bar\Delta}&4\xi_{\vec k}+2U
 \end{array}
 \right)
\eeq
with corresponding 3 energy levels ($i=1,2,3$)
\beq
 E_i=2 \xi_{\mathbf{k}}+\frac{2 U}{3}+\frac{4}{\sqrt{3}} E_{\mathbf{k}}^{\text {even }} \cos \left(\theta_{\mathbf{k}}+\frac{2 \pi}{3} i\right)\\
 E_{\mathbf{k}}^{\mathrm{even}}=\sqrt{\left(\xi_{\mathbf{k}}+\frac{U}{2}\right)^{2}+\bar\Delta^{2}+\frac{U^{2}}{12}}\\
 \theta_{\mathbf{k}}=\frac{1}{3} \arccos \left[\frac{q_{\mathbf{k}}}{\left(\sqrt{3} E_{\mathbf{k}}^{\text {even }}\right)^{3}}\right]\\
 q_{\mathbf{k}}=U\left(U^{2}+\frac{9}{2} U \xi_{\mathbf{k}}+\frac{9}{2} \xi_{\mathbf{k}}^{2}-\frac{9}{4} \bar\Delta^{2}\right).
\eeq
Similarly, the odd sector can be written into a direct-sum of 4 equivalent subspace, e.g. \{$\ket{1000}, \ket{1110}$\}. Here the Hamiltonian matrix in each 2-dimensional subspace is
\beq
 \left(
 \begin{array}{cc}
  \xi_{\vec k}&-\bar\Delta\\
  -\bar\Delta&3\xi_{\vec k}+U
 \end{array}
 \right)
\eeq
and the odd sectors share the 2 energy levels,
\beq
 E_{\pm}=2 \xi_{\mathbf{k}}+\frac{U}{2} \pm E_{\mathbf{k}}^{\text {odd }}\\
 E_{\mathbf{k}}^{\text {odd }}=\sqrt{\left(\xi_{\mathbf{k}}+\frac{U}{2}\right)^{2}+\bar\Delta^{2}}.
\eeq
The energy levels and bases in each subspace are shown in Table \ref{tab:energylevels}.  Thus we have all energy levels of the system. The ground state is recognized as the eigenvector coresponding to $E_1$, which is a linear combination of the 3 occupancy states: $\ket{0000}$ from $\Omega_0\otimes\Omega_0$, $(\ket{1001}+\ket{0110})/\sqrt2$ from $\Omega_1\otimes\Omega_1$ and $\ket{1111}$ from $\Omega_2\otimes\Omega_2$. Our previous variational treatment\cite{hksupercon} concurs with this result.

The work-horse of the mean-field treatment is the self-consistent equation for the pair amplitude: 
\beq
 \bar{\Delta}=\frac{g}{V}\sum_{\vec k}\ave{c_{\vec k\uparrow}c_{\vec{-k}\downarrow}}.
\eeq
The right-hand side has a dependence on $\bar\Delta$ through the thermal average over the energy levels enumerated in Table \ref{tab:energylevels}. We can also treat the self-consistent equation for $\bar\Delta$ as the solution to the extremum of the free energy,
\beq
F=-k_BT\ln Z,
\eeq
where
\beq
Z=\sum_{i,k\in HFBZ} e^{-\beta E_{\vec k}^i}.
\eeq
\begin{figure}[ht]
 \centering
 \includegraphics[width=8.6cm]{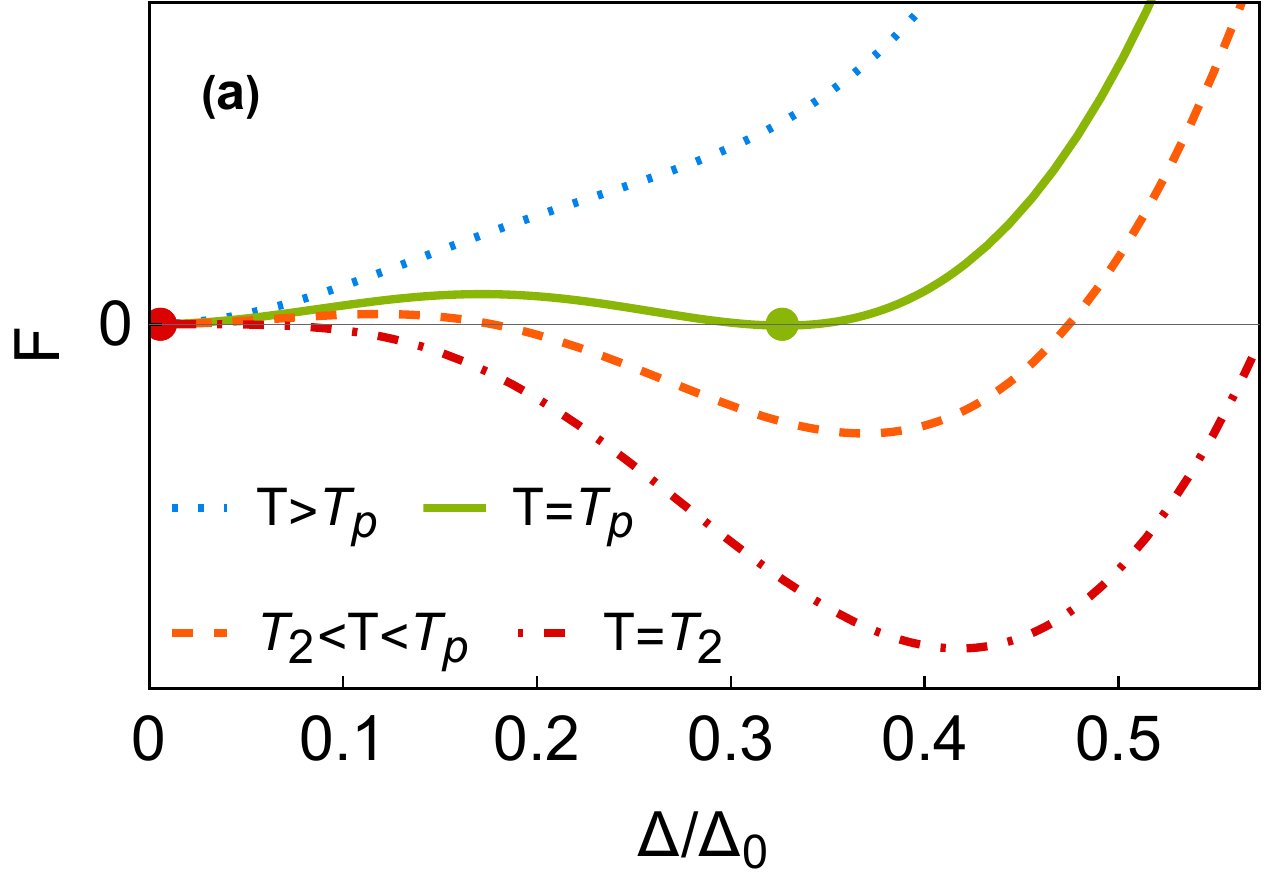}
 \includegraphics[width=8.6cm]{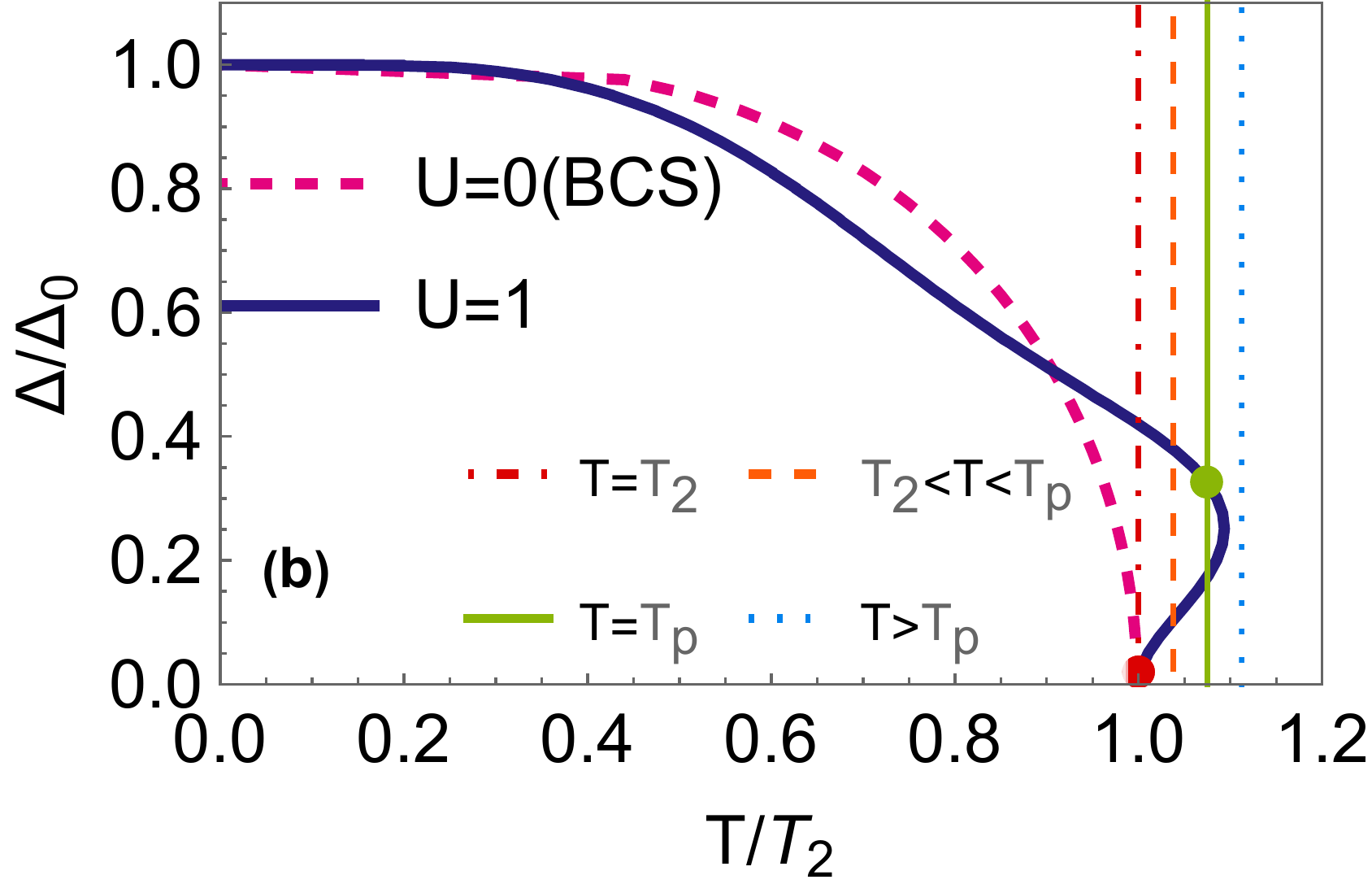}
 \caption{(\textbf{a}) The variation of the free Energy $F$ as a function of the the pair amplitude $\Delta$ for an HK superconductor at $U/W=1$ and $g=0.3$. The chemical potential is set at half filling of the lower band ($\mu=0.5$). For $T>T_p$, the global minimum of free energy is $\Delta=0$. At $T=T_p$, the free energy at $\Delta=0$ and $\Delta=\Delta_p$ coincides, thus indicating a first-order transition. When $T_2<T<T_p$, $\Delta=0$ is still a local minimum, while the global minimum obtains at finite $\Delta$. For $T= T_2$, the only minimum occurs at finite $\Delta$, and $\Delta=0$ becomes a local maximum(the inflection point).
 (\textbf{b}) The dashed purple line and the solid blue line are the solution to the pair amplitude from the self consistent equation at different temperatures(the $\Delta-T$ plot) for BCS($U/W=0$) and HK($U/W=1$). The Green dot shall represent the first-order phase transition and the red dot represents the second-order phase transition. In both cases, the axes are scaled according to respective $T_2$, and respective zero temperature pair amplitude $\Delta_0$.
 }
 \label{fig:pairamplitude}
\end{figure}

Both procedures yield identical results and are tabulated in Fig. \ref{fig:pairamplitude}. Displayed first is the free energy as $\bar\Delta$ is varied. Of first note is that for $T>T_p$,
the global minimum of free energy remains zero. At $T=T_p$ ($p$ for pair), there are two degenerate minima, with $\bar\Delta$=0 and $\bar\Delta=\Delta_p\ne 0$. The technical definition of $T_p$ then is $\partial F/\partial\Delta|_{\bar\Delta=\Delta_p}=0$ and $F(\Delta_p)=F(0)$. This degeneracy is lifted by lowering the temperature such that $T_2<T<T_p$. The solution to $\partial^2 F/\partial\Delta^2|_{\Delta=0}=0$ defines $T_2$ as the inflection point. As we will see, $T_2$ will correspond to the divergence of the susceptibility.

To corroborate this picture, we solve the self-consistent equations for the the gap. The dashed curve in Fig. \ref{fig:pairamplitude}(b) corresponds to BCS theory which has a unique turn-on temperature for the pair amplitude at the red dot, namely $T=T_2$. In HK, however from our evaluation of the free energy, we see multiple solutions for the gap turn-on distinct from the feature at $T_2$. As shown, the $\Delta-T$ curve for the mean-field HK exhibits a nontrivial back-folding behavior above $T_2$. This multivaluedness confirms that there are multiple choices of $\Delta$ that make the first derivative of the free energy with respect to the gap vanish. Only one of these gaps yields the global minimum of the free energy. The equivalence of the free energy at $\Delta=0$ and $\Delta_p$ implies a first order phase transition at this temperature $T_p$(green dot), which ensures the accuracy of mean-field treatment since the first order transition does not have any diverging fluctuations.   The degenerate minimum of the free energy at $T=T_p$ are shown in Fig. \ref{fig:pairamplitude}(a) with the green solid curve. There are two mechanisms to lift the degeneracy: 1) decrease the temperature as shown in Fig. (\ref{fig:pairamplitude}a) or 2) decrease $U$ from the value shown in Fig. (\ref{fig:pairamplitude}b). What is not shown is that there is a critical value of $U$ that is needed to destroy the back-folding in Fig. (\ref{fig:pairamplitude}b). This critical value diminishes as the pairing strength, $g$, decreases. While it is suggestive at this point that it is Mott physics that leads to the first-order nature of the superconducting transition, we will confirm this by a detailed evaluation of the Landau expansion parameters.
\begin{figure}
    \centering
    \includegraphics{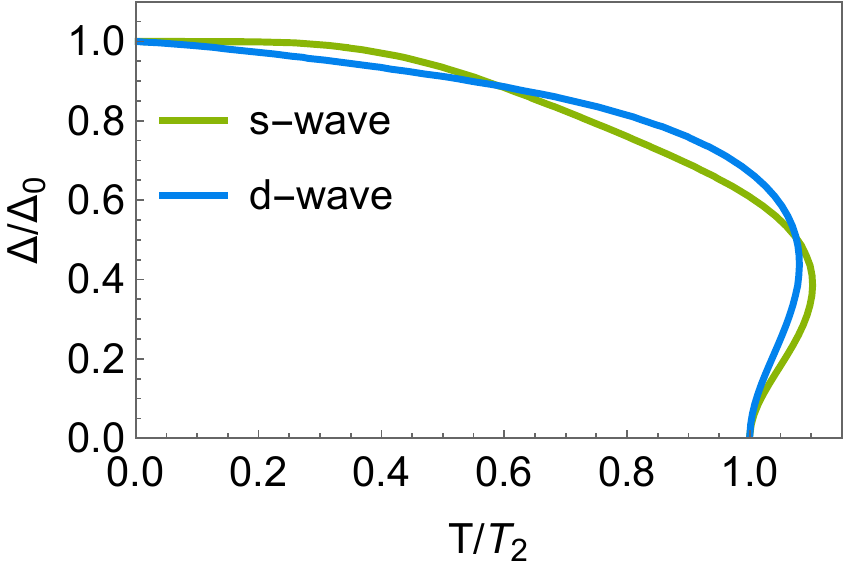}
    \caption{Pair amplitude of s-wave and d-wave pairing at the same pairing strength $g=0.3$ and repulsion strength $U/W=1$.}
    \label{fig:sd-wavepa}
\end{figure}
An analysis of the HK model under d-wave pairing could be carried out similarly by replacing the pairing term $H_p$ in Eq.(\ref{Htot}) with a d-wave form
\beq
H_d=\frac{g_d}{V} \sum_{\vec k,\vec k'}\tilde{b}_{\vec k}\adj \tilde{b}_{\vec k'},
\eeq
where $\tilde{b}_{\vec k}= c_{-\vec k\downarrow} c_{\vec k\uparrow}(\cos k_x-\cos k_y)$ is the $d$-wave pair creation operator in 2-dimensions. After performing the same mean-field calculation on $\tilde{b}_{\vec k}$, we find that the pair amplitude, as shown in Fig. \ref{fig:sd-wavepa}, also exhibits a nontrivial back-folding behavior above $T_2$, which is recognized as the key major difference between a BCS superconductor and an HK superconductor at the mean field level.   As there is little difference between these two cases, in the subsequent analysis only the s-wave case will be analyzed.

One remaining subtlety is that the calculations in this section are all based on mean-field theory, which is exact only if the transition is truly first order.  Strictly, they break down at the critical point of a second order transition. Since the exact pair susceptibility diverges at a temperature distinct from $T_p$, we need to consider two possibilities: 
\begin{enumerate}
    \item The phase transition is first-order at $T_p$, and the calculation of the susceptibility cannot be applied to the theory below the transition temperature at which $T_p$($\Delta$ jumps to a finite value instead of continuously growing from zero when $T=T_p$). The mean-field calculation is exact in this scenario.
    \item The phase transition is second-order at $T_c$. The global minimum of the free energy at $T_p$ shall be excluded due to the divergence of fluctuations near the second-order phase transition, which destroys the mean-field theory. Only the non-MF calculation (e.g. susceptibility) is right.
\end{enumerate}
In the first case, the calculation of $T_c$ is internally inconsistent as it presumed that $\Delta_p=0$ for $T>T_c$.   In the second case, we take advantage of the exact diagonalizability and use the Ginzburg criterion to estimate the temperature range where the fluctuations diverge (mean-field theory breaks down).

\subsection{Ginzburg Criterion}

The unusual first-order nature of the phase transition in an HK superconductor is manifest at the mean-field level. The crucial question is then: Is this first-order Transition just an artifact? In other words, how valid is the mean-field theory, or equivalently, what is its range of validity? To address this, we compute the Ginzburg criterion, presuming that the phase transition is second-order at $T=T_c$. 

Recall that the Landau-Ginzburg description of a traditional superconductor is equivalent to the mean-field treatment of the BCS theory around the transition point\cite{coleman2015introduction}. We demostrate the same result for HK superconductors. 
In an s-wave superconductor which is the case considered here, the order parameter is given by
\begin{equation}
 \Psi(x)\propto\ave{c_\uparrow(x)c_\downarrow(x)}.
\end{equation}
In the absence of a magnetic field, the Landau free energy is 
\begin{equation}
 F[\Psi]=\int d^dx\left(a|\Psi(x)|^2+\frac{1}{2}b|\Psi(x)|^4+K|\nabla\Psi(x)|^2\right),
\end{equation}
where $a$ is directly related to the susceptibility and $K$ is the rigidity. 
Close to a second-order critical point $T_c$, that is $|\Psi|\ll1$, we may neglect the quartic terms. As a result, the free energy,
\begin{equation}
\begin{split}
 F[\Psi,\Psi^*]&=\int d^dx\left(a|\Psi(x)|^2+K|\nabla\Psi(x)|^2\right)\\
 &=\sum_{\vec k}(a+Kk^2)|\Psi(k)|^2\\
 \Psi(x)&=\frac{1}{\sqrt{V}}\sum_{\vec k}\Psi(k)e^{ikx},
\end{split}
\end{equation}
by performing the Fourier transform of the order parameter.
The Helmholtz free energy $A(T)$ is given by
\begin{equation}
\begin{split}
 A(T)&=-k_BT\ln Z\\
 &=-k_BT\ln\int D[\Psi,\Psi^*]e^{-F[\Psi,\Psi^*]/k_BT}\\
 &=k_BT\sum_{\vec k}\ln\left(\frac{k_BT}{a+Kk^2}\right).
\end{split}
\end{equation}
We write $a(T)=\alpha t$ with $t=\frac{T-T_c}{T_c}$. The singular contribution to the specific heat $C_V=-TA''(T)$ comes when differentiating with respect to $T$. The dimensionless heat capacity per unit cell is
\begin{equation}
\begin{split}
 c&=\frac{C_V}{N_sk_B}\\
 &=\frac{\alpha^2 \mathsf{a}^d}{K^2}\int^{\Lambda}\frac{d^dk}{(2\pi)^d}\frac{1}{(\xi^{-2}+k^2)^2}\\
 &=\frac{\alpha^2 \mathsf{a}^d}{K^2}\xi^{4-d}\int^{\Lambda\xi}\frac{d^dq}{(2\pi)^d}\frac{1}{(1+q^2)^2},
\end{split}
\end{equation}
where $\Lambda\sim\mathsf{a}^{-1}$ is an ultravilot cut-off and $\mathsf{a}$ is the lattice constant(set to unit in the following calculation), $\xi=(K/a)^{1/2}=(K/\alpha)^{1/2}|t|^{-1/2}$. The fluctuation contribution is small when 
\begin{equation}
 \frac{\alpha^2 \mathsf{a}^d}{K^2}\xi^{4-d}\ll1,
\end{equation}
which entails $|t|\gg t_G$ where the latter is defined,
\begin{equation}
 t_G=\left(\mathsf{a}^2\frac{\alpha}{K}\right)^{\frac{d}{4-d}},
 \label{eq:ginzburgt}
\end{equation}
as the Ginzburg reduced temperature.
This calculation then just hinges on knowing $\alpha$ and $K$ and we can estimate the temperature region where mean-field theory is not valid.

\begin{figure}[ht]
 \centering
 \includegraphics[width=8.6cm]{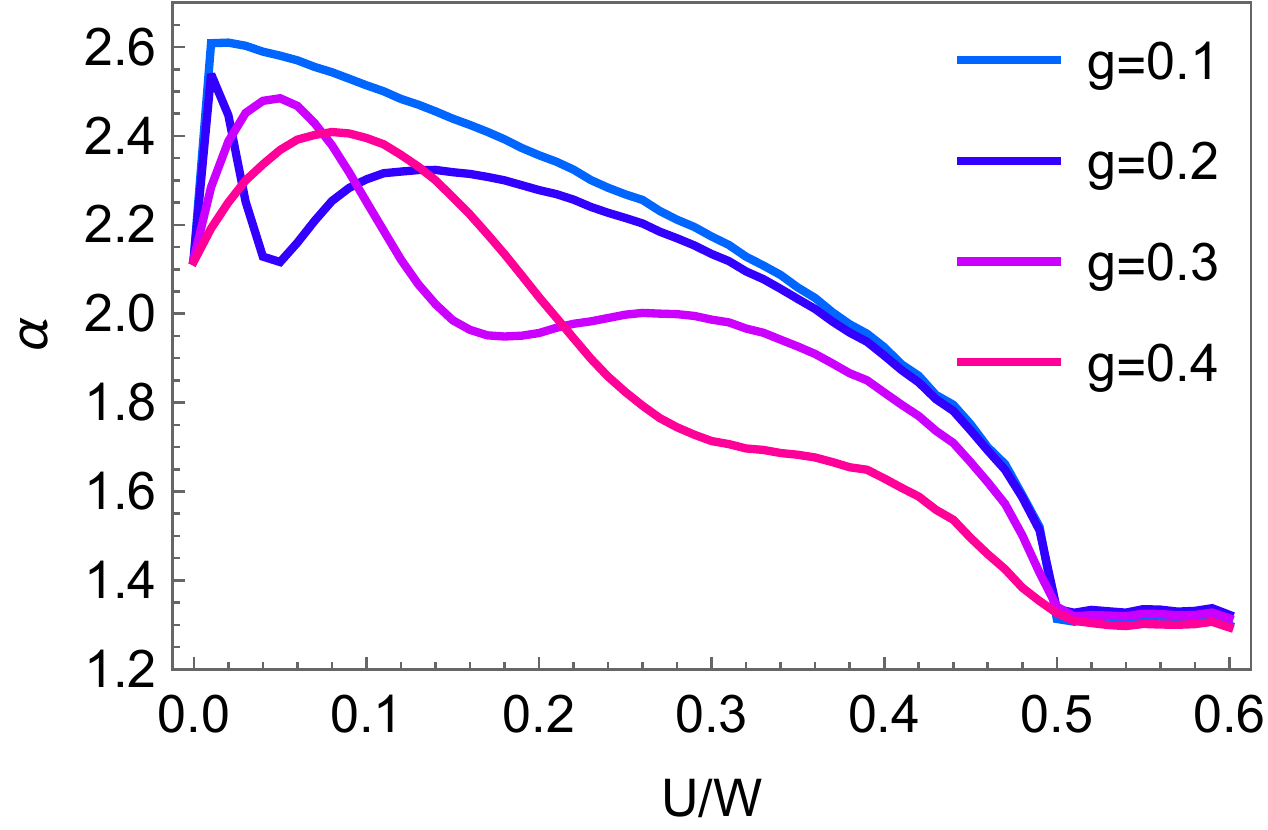}
 \caption{The dependence of the Landau expansion parameter $\alpha=a/t$ on the HK interaction strength $U$ at different values of the superconducting pairing strength $g$. We claim that at $U=0$, $\alpha\approx2.1$ and at $U>\mu=0.5$, $\alpha\approx1.3$ which are around the same scale. The magnitude of $\alpha$ does not vary appreciably for all values of $g$.}
 \label{fig:ginzburgalpha}
\end{figure}

To obtain $\alpha$,
we calculate the spatial average of the order parameter
\begin{equation}
\begin{split}
 \bar{\Psi}&\equiv\frac{g}{V}\int d^dx \ave{c_\uparrow(x)c_\downarrow(x)}\\
 &=\frac{g}{V}\sum_{\vec k}\ave{c_{\vec k\uparrow}c_{\vec{-k}\downarrow}}\\
 &=\bar{\Delta}.
\end{split}
\end{equation}
Thus, the Landau free energy parameters can be calculated by differentiating the free energy density $f$ with respect to $\bar\Delta^2$
\begin{equation}
 a=\left.\frac{\partial f}{\partial \bar\Delta^2}\right|_{\bar\Delta=0},
\end{equation}
and the parameter $\alpha$ can be calculated by linear fitting the $a(T)$ around the transition temperature as shown in Fig. \ref{fig:ginzburgalpha}. The key point is that $\alpha$ does not vary appreciably for the system parameters $U/W$ and $g$.

\begin{figure}[ht]
 \centering
 \includegraphics[width=8.6cm]{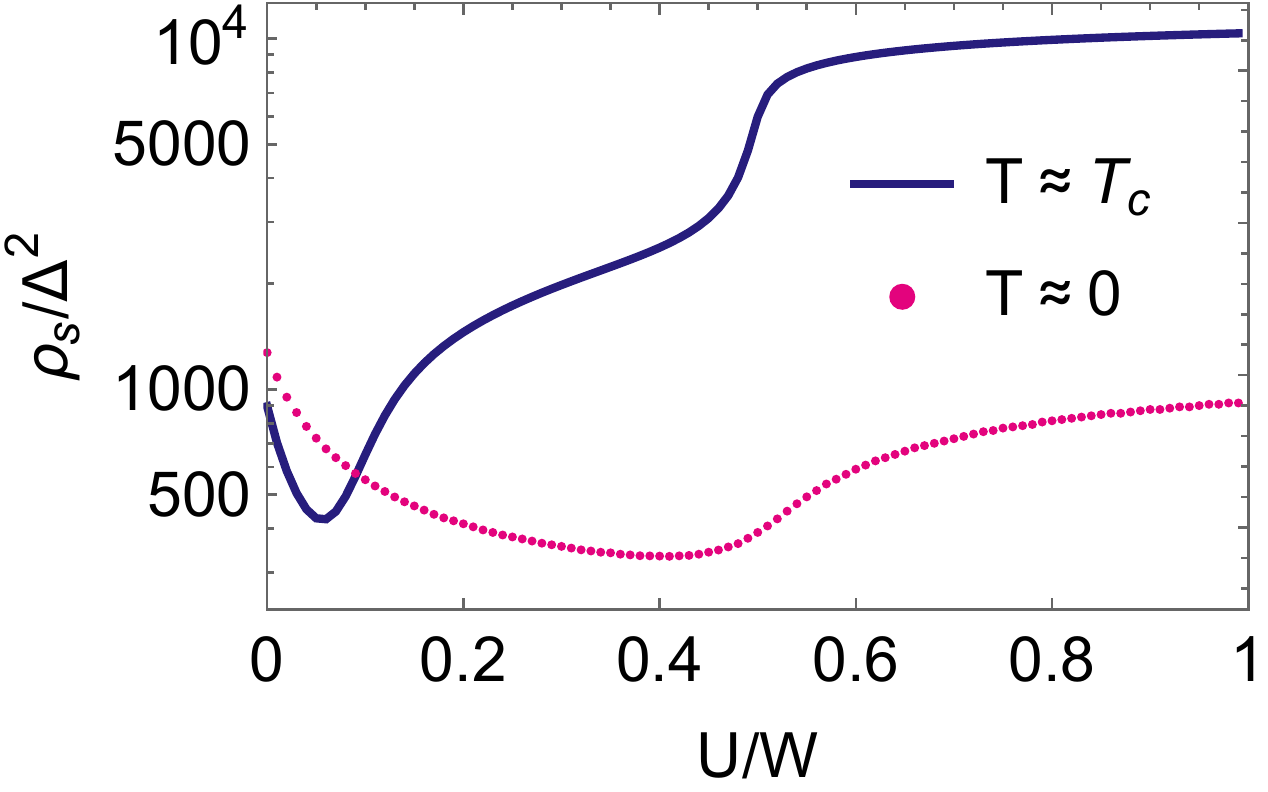}
 \caption{The rigidity at transition temperature($T\approx T_c$, solid line) and at zero temperature($T\approx0$, dotted points). The calculation(see appendix) was performed with $\mu=0.5$, $g=0.2$. The value of $K$ was calculated with $\bar\Delta=10^{-5}$ and $T=T_c$(where the susceptibility diverges).}
 \label{fig:stiffness}
\end{figure}

We also have the superfluid stiffness defined by
\begin{equation}
 T[\theta]=\int d^dx\frac{1}{2}\rho_s(\nabla\theta)^2,
\end{equation}
where $\theta$ is the phase of the superfluid $\Psi(x)=\bar\Psi e^{i\theta}$. Thus the relation between $K$ and $\rho_s$ is 
\beq
 \frac{1}{2}\rho_s(\nabla\theta)^2=K|\nabla\Psi|^2=K\bar\Psi^2(\nabla\theta)^2\\
 K=\frac{\rho_s}{2\bar\Psi^2}=\frac{\rho_s}{2\bar\Delta^2}.
\eeq
From Fig. \ref{fig:ginzburgalpha} and Fig. \ref{fig:stiffness}, we can read that at $\mu=0.5$, $U>\mu$, the value of $\alpha$ and $\rho_s$ is $\alpha\approx1.3$, $ K\approx\rho_s/ 2\Delta^2\approx10^4$. From Eq. (\ref{eq:ginzburgt}), we may estimate $t_G\approx10^{-11}$. Consequently, the Ginzburg temperature is sufficiently small to guarantee the validity of the mean-field calculation, which predicts a first order transition. Hence, the presumption that the phase transition is second-order shall be ruled out when $T_p>T_2$. We will discuss the consequences of this first order transition in more depth.


\subsection{Phase Diagram}

\begin{figure}[ht]
 \centering
 \includegraphics[width=8.6cm]{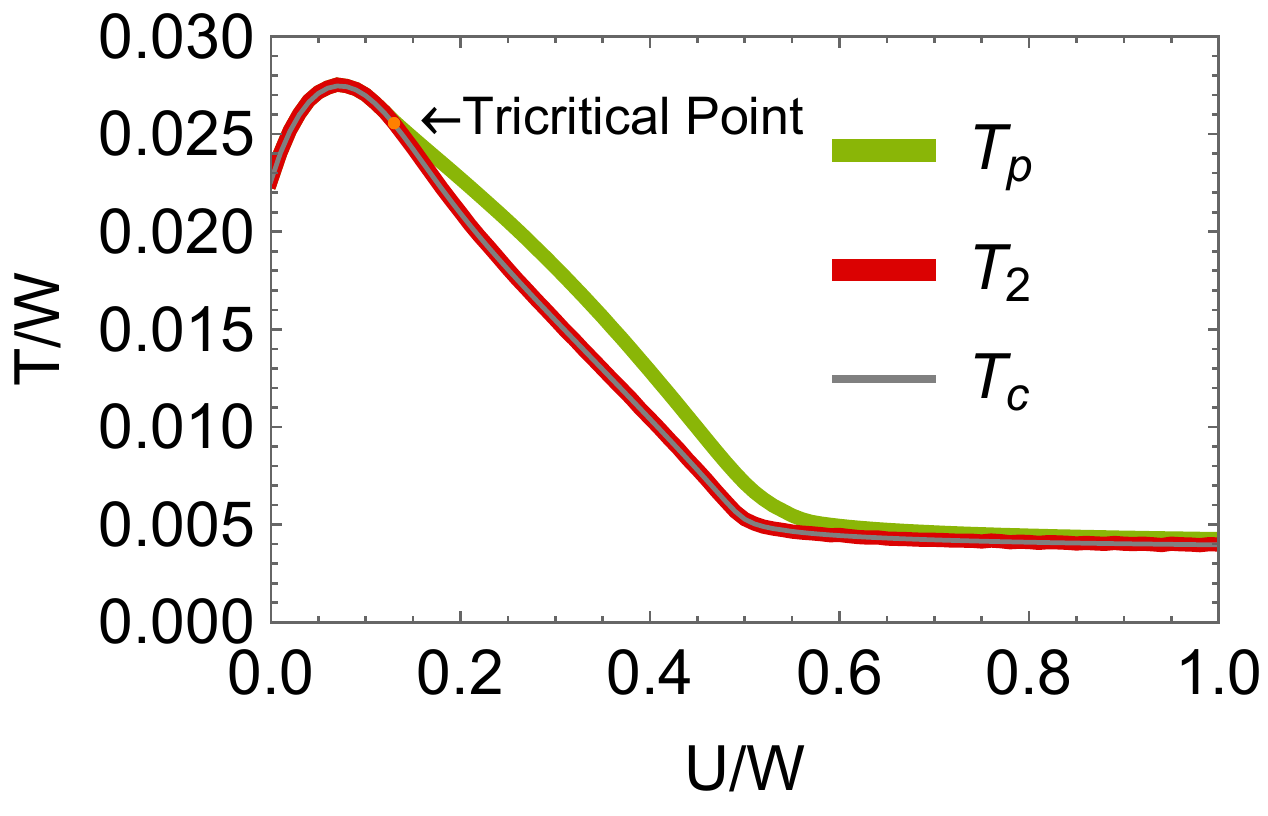}
 \caption{Phase diagram in the $T-U$ plane of the HK superconductor for coupling $g=0.3$.The green line corresponds to the first-order transition temperature $T_p$ at which the global minimum of the free energy switches from $\Delta=0$ to a finite value. The red line corresponds to $T_2$ at which $\Delta=0$ changes from a local minimum into a local maximum. The gray line is the pair susceptibility diverging temperature $T_c$. The gray and red line coincide, thereby corroborating that the pair-susceptibility divergence is coincident with the inflection point in the free energy.}
 \label{fig:phasediagram}
\end{figure}

To put this all together, we focus on the 
phase diagram in the $T-U$ plane presented in Fig. \ref{fig:phasediagram} with a superconducting paring strength $g=0.3$.  We observe that for a small value of $U/W<0.13$, the superconducting transition is second-order (red line) in which the divergence of the pair susceptibility and the turn-on of the gap are coincident. In general, the critical value of $U/W$, denoted by $U_t/W$, for the transition to be second order decreases as $g$ decreases. Hence, it is only the weakly interacting regime in which we find the traditional result that the superonducting transition is second order.  As pointed out in our previous paper, in no regime (except $U=0$) does a BCS picture apply.   For example, even in the weakly interacting regime, the  transition temperature compared with BCS($U=0$) value increases as $U$ increases implying that the multiband nature of the HK model is at play in driving the enhancement of superconductivity.  The second-order transition line ends at the tricritical point located at $U_t/W=0.13$ and $T_t/W=0.0255$.   It is at this point that the local and global minima merge.  For larger values of $U>U_t$, the transition becomes a first-order one(green dots). The phase transition temperature $T_p$ saturates when $U/W>0.5$ which coincides the elimination of the single-occupancy region $\Omega_1$ and double occupancy region $\Omega_2$ boundary in the HK model. 

We can also compare $T_2$ with $T_c$. As is evident, they both coincide as illustrated in Fig. \ref{fig:phasediagram}. This is significant because $T_c$ is computed from a divergence of the susceptibility while $T_2$ follows from a solution to a mean-field equation.
For an HK superconductor with first order transition, the temperature $T_2$ represents the elimination of the metastable state(local minimum). 

\begin{figure}[ht]
 \centering
 \includegraphics[width=8.6cm]{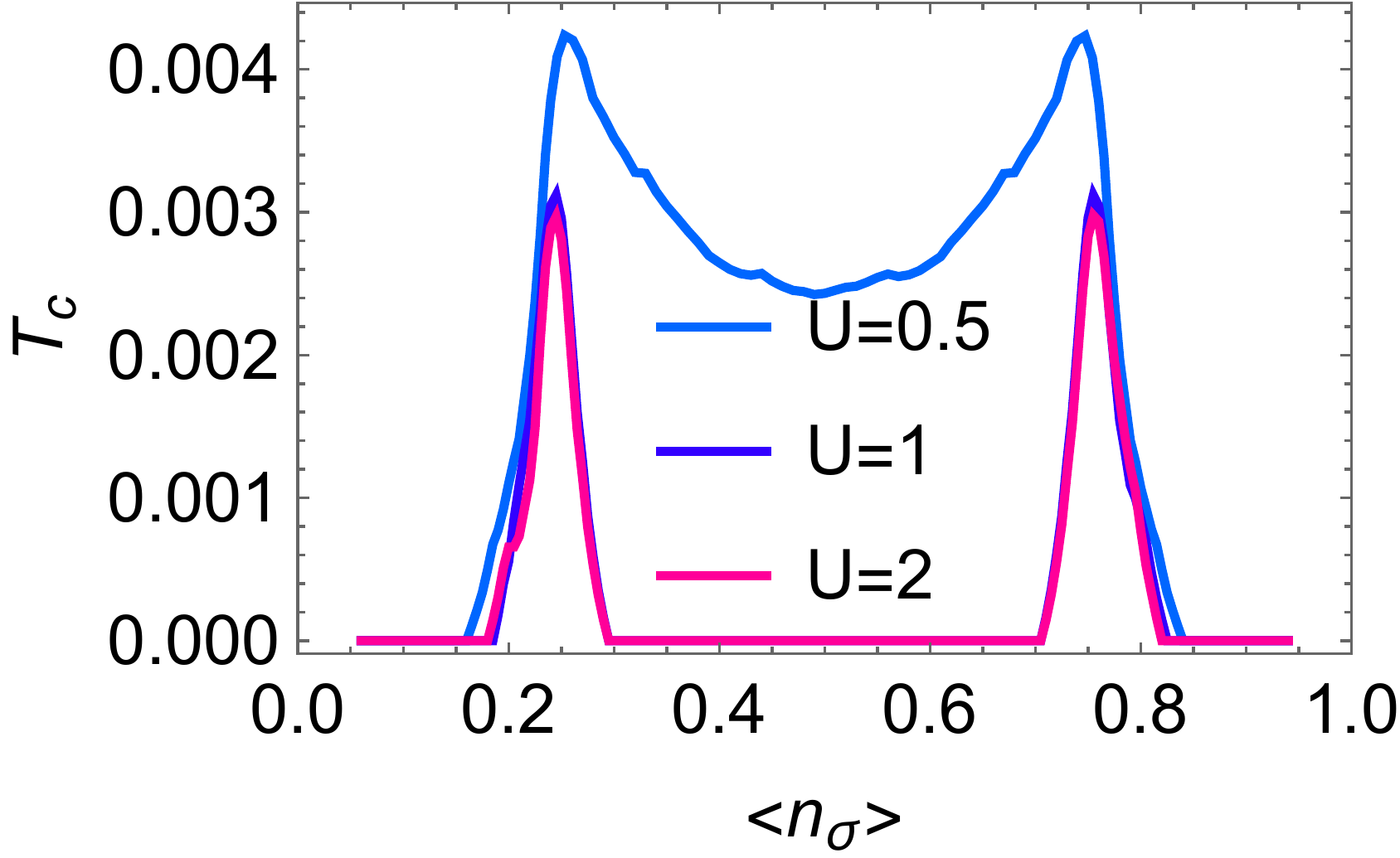}
 \caption{Phase diagram in the $T_c-\ave{n_\sigma}$ plane of the HK superconductor for coupling $g=0.3$. The blue line ($U=0.5$) represents the overlapping bands or metallic phase in which $U<W$. The pink and purple line($U=1,2$) represent the non-overlapping bands (Mott insulator ground state at half filling, $\ave{n_\sigma}=0.5$).}
 \label{fig:mudiagram}
\end{figure}

In all the computations thus far, either the chemical potential or the filling has remained fixed.  To make contact with the cuprates, a phase diagram of $T_c$ versus filling or chemical potential is needed.  This can be done using the same machinery.  Shown in Fig. \ref{fig:mudiagram} is a plot of $T_c$ versus filling for two distinct cases: 1) overlapping lower and upper Hubbard bands and 2) no overlap.  This can be engineered simply by changing the value of $U$.  In general electron and hole doping yield the mirror results.  When the bands overlap, a metallic state always ensues, though a non-Fermi liquid one\cite{hksupercon} and $T_c$ is minimized at half-filling though it does not vanish.  A vanishing of $T_c$ at the Mott insulating state obtains only for $U>W$ as is shown in the plots in which $U=1$ and $U=2$.  The familiar dome-shaped phase diagram obtains as dictated by the dome-shaped superfluid stiffness reported previously\cite{hksupercon}.    The dome shape here is a direct consequence of Mott physics. 

\subsection{Landau Expansion Parameters}

\begin{figure}[ht]
 \centering
 \includegraphics[width=8.6cm]{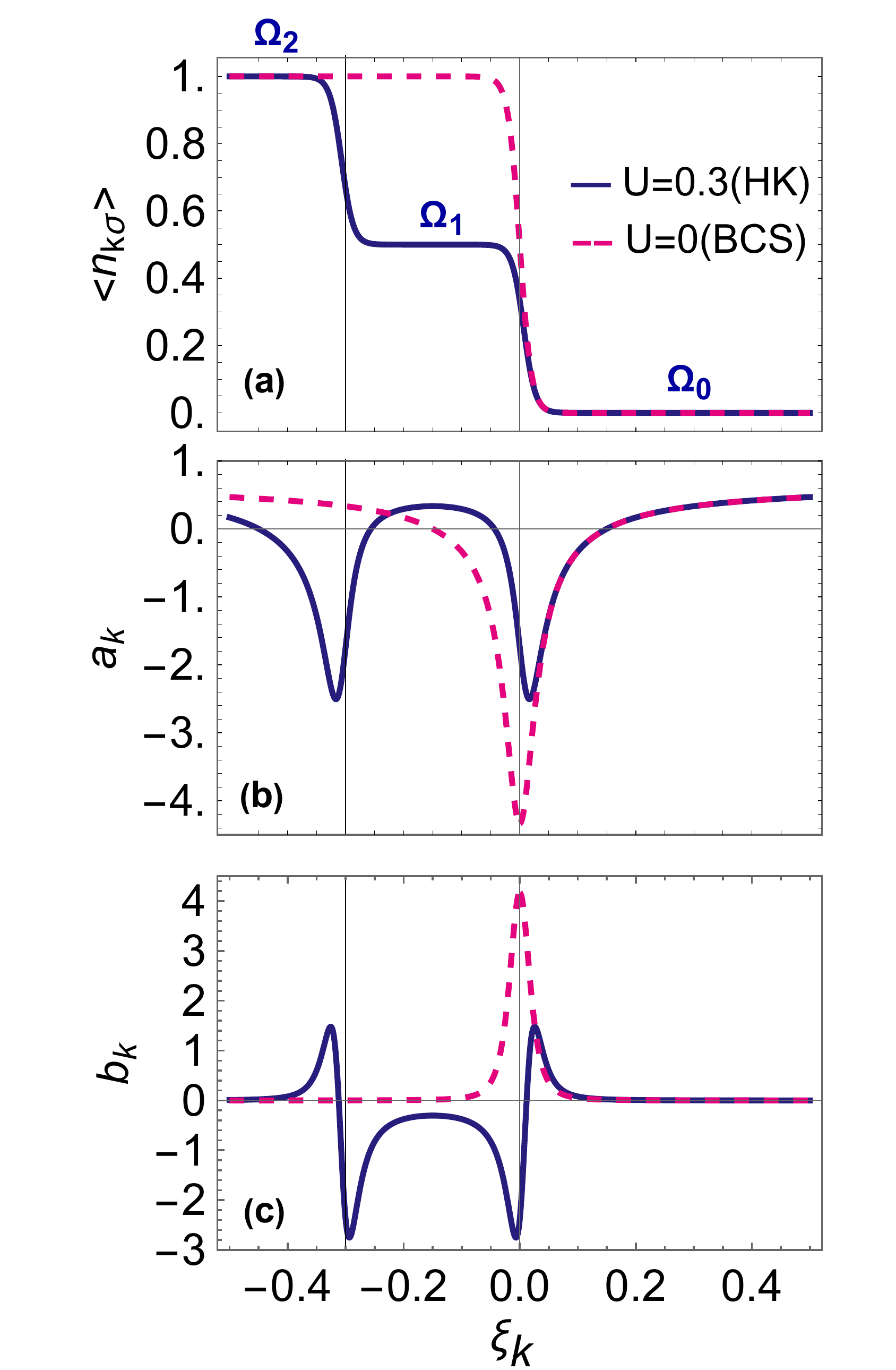}
 \caption{(a) The occupancy of the HK(U=0.3) and Fermi liquid(FL) at finite temperature $T/W=0.01$. 
 (b) The 2nd expansion parameter $a_{\vec k}$ of the Landau Free Energy over the energy levels $\xi_{\vec k}$. 
 (c) The 4th expansion parameter $b_{\vec k}$ of the Landau Free Energy over the energy levels $\xi_{\vec k}$.}
 \label{fig:landauparameters}
\end{figure}

To complete a theory of a first order transiton, we rewrite the Landau free energy
\beq
 F[\bar\Delta]=a\bar\Delta^2+b\bar\Delta^4+c\bar\Delta^6+O(\bar\Delta^6).
\eeq
in terms of the 4th order expansion parameter $b$. For a first-order transition, $b<0$ and the 6th order term be positive to keep the free energy bounded from below. When the first-order transition obtains, both the value of the free energy and the first derivative with respect to $\delta$ vanish,
\beq
 F[\bar\Delta_p]&=a\bar\Delta_p^2+b\bar\Delta_p^4+c\bar\Delta_p^6=0\\
 F'[\bar\Delta_p]&=2a\bar\Delta_p+4b\bar\Delta_p^3+6c\bar\Delta_p^5=0.
\eeq
The solution is $\bar\Delta_p^2=-2a/b$ and $b^2=4ac$, thus proving that $a>0, b=-\sqrt{4ac}<0$ is required for this kind of first order transition to exist. Recall that for an HK superconductor, extreme factorizability in momentum sections allows us to write the free energy as
\beq
 F[\bar\Delta]=\sum_{k\in HFBZ}F_{\vec k}[\bar\Delta].
\eeq
As a result, the free-energy expansion parameters can be decomposed as 
$a=\sum_{\vec k}a_{\vec k}$ and $b=\sum_{\vec k}b_{\vec k}$. To organize our results, we recall that what makes the HK model a non-Fermi liquid is that at zero-temperature, as a result of $U$, singly occupied states, $\Omega_1$, exist below the chemical potential which never obtains in a Fermi liquid. This is depicted in Fig. \ref{fig:landauparameters}(a). The results for the momentum-resolved Landau expansion coefficients are shown in panels Fig. \ref{fig:landauparameters}(b,c). 

In the BCS case, the negative contribution of $a_{\vec k}$ comes from the states around the Fermi surface, where the occupancy changes from 2 to zero. This arises from the sharpness of the Fermi surface. In the HK superconductor, similar sign changes at the occupancy boundaries from $\Omega_2$ to $\Omega_1$ and from $\Omega_1$ to $\Omega_0$. $a_{\vec k}$ is always possitive except around the boundary of different occupancy region. When the positive contribution exactly cancels, the negative contribution, $a$ vanishes, implying that the free energy changes from a concave function into a convex one around $\Delta=0$. Most crucial here, however, is the behaviour of $b_{\vec k}$. 
For the HK superconductors, while the distribution of $a_{\vec k}$ follows the BCS case, the value of $b_{\vec k}$ differs drastically from BCS. In BCS, $b_{\vec k}$ is always positive; thus $b>0$ is true for any temperature. In HK, however, inside the single occupancy region $\Omega_1$, which is not present in any fermi liquid, $b_{\vec k}$ becomes negative! Together with a surpression of the positive $b_{\vec k}$ contribution, it is possible that $b<0$ and a first order transition emerges. Consequently, it is Mottness that drives the first-order nature of the transition in the HK model. Perhaps this is true in general.

\subsection{Condensation Energy}

\begin{figure}[ht]
 \centering
 \includegraphics[width=8.6cm]{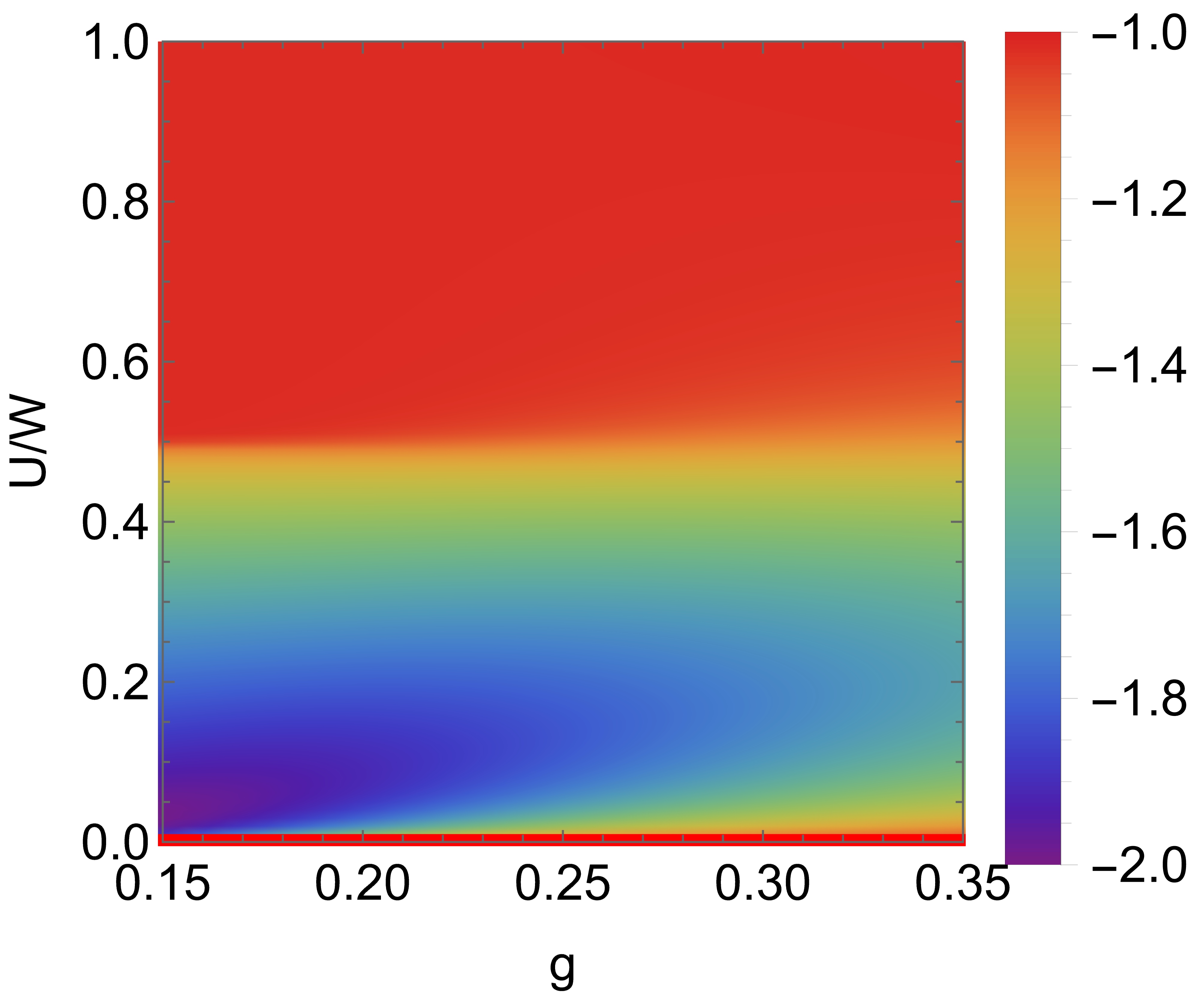}
 \caption{The dependence of the condensation energy $E_{\rm cond}$ devided by $N(0)\Delta^2$ at zero temperature in the $g-U$ plane. The chemical potential is set $\mu/W=0.5$. The bottom red bold line represents the BCS result $C=-1$. As is evident, HK superconductors have condensation energies that generically deviate from the BCS result. }
 \label{fig:condensation}
\end{figure}

The deviation of an HK superconductor from that of the BCS type is also manifest from the condensation energy, defined by the energy difference between the superconducting and normal states,
\beq
E_{\rm cond}=E_{SC}-E_N.
\eeq
In a traditional s-wave BCS superconductor, the condensation energy is well-known to be propotional to the square of the pair amplitude $E_{\rm cond}^{BCS}=-N(0)\Delta^2/2$, where $N(0)$ is the density of states around the chemical potential. In the HK superconductor, however, this relation holds up to a modification to the coefficient 
\beq
E_{\rm cond}^{HK}=CN(0)\Delta^2/2,
\eeq
where $C$ is a pure number. The dependence of $C$ on pairing strength $g$ and $U$ is plotted in Fig. (\ref{fig:condensation}).
Generally, the coefficient $C$ have a smaller value in HK superconductors than in BCS, demonstrates a stronger tendency to form a superconductor. This result is surprising but consistent with what we know about the cuprates in which the pairing temperatures are greatly enhanced despite the local repulsions. In this case, enhancement arises because of the singly-occupied region. The minimum value $C\approx-2$ can be achieved at a specific point $U$ smaller than the chemical potential and quickly goes back to the BCS ratio $C=-1$ as $U>\mu$, where the double occupancy region $\Omega_2$ goes away, and the boundary between $\Omega_1$ and $\Omega_2$ is removed. This observation again emphasizes the significance of the single occupancy region boundary in the HK superconductor. It may imply a mechanism to enhance the superconductivity by adding the occupancy boundary, where the Copper pair formation is enhanced. 

\subsection{Heat capacity}

Even when the superconducting gap develops continuously at the BCS transition, the heat capacity $C(T)$ undergoes a jump discontinuity $\Delta C$ with the universal ratio $\Delta C / C_n(T_c) = 12/7\zeta(3) \approx 1.43$, where $C_n(T_c)$ is the heat capacity in the normal phase at $T=T_c$.
The discontinuity obtains because the BCS Hamiltonian is an effective one that varies with the temperature through the gap parameter $\Delta(T)$.  The same is true of HK even with a first-order transition as can be seen from the simple argument.
With the free energy given by $-T\ln \tr e^{-H/T} = \ave{H} - TS$, any system with a temperature-independent Hamiltonian has a heat capacity coefficient $C/T \equiv \partial_T S = \beta^3 \ave{(H - \ave{H})^2}$ that varies continuously with the spectrum of $H$.
However, because the gap is temperature dependent, the heat capacity
\begin{equation}
    C/T
    = \beta^3 \B{ \ave{(H - \ave{H})^2} - \Delta \pd{\Delta}{T} \ave{H} },
\end{equation}
has an explicit temperature derivative.  This term introduces a jump discontinuity at $T = T_c$ since $\Delta(T) \sim \sqrt{T_c-T}$ (so $\Delta \partial_T \Delta \sim -1$) on the superconducting side of the transition whereas $\Delta(T) = 0$ (so $\Delta \partial_T \Delta = 0$) on the normal side.

\begin{figure}[htpb]
 \centering
 \includegraphics[width=\linewidth]{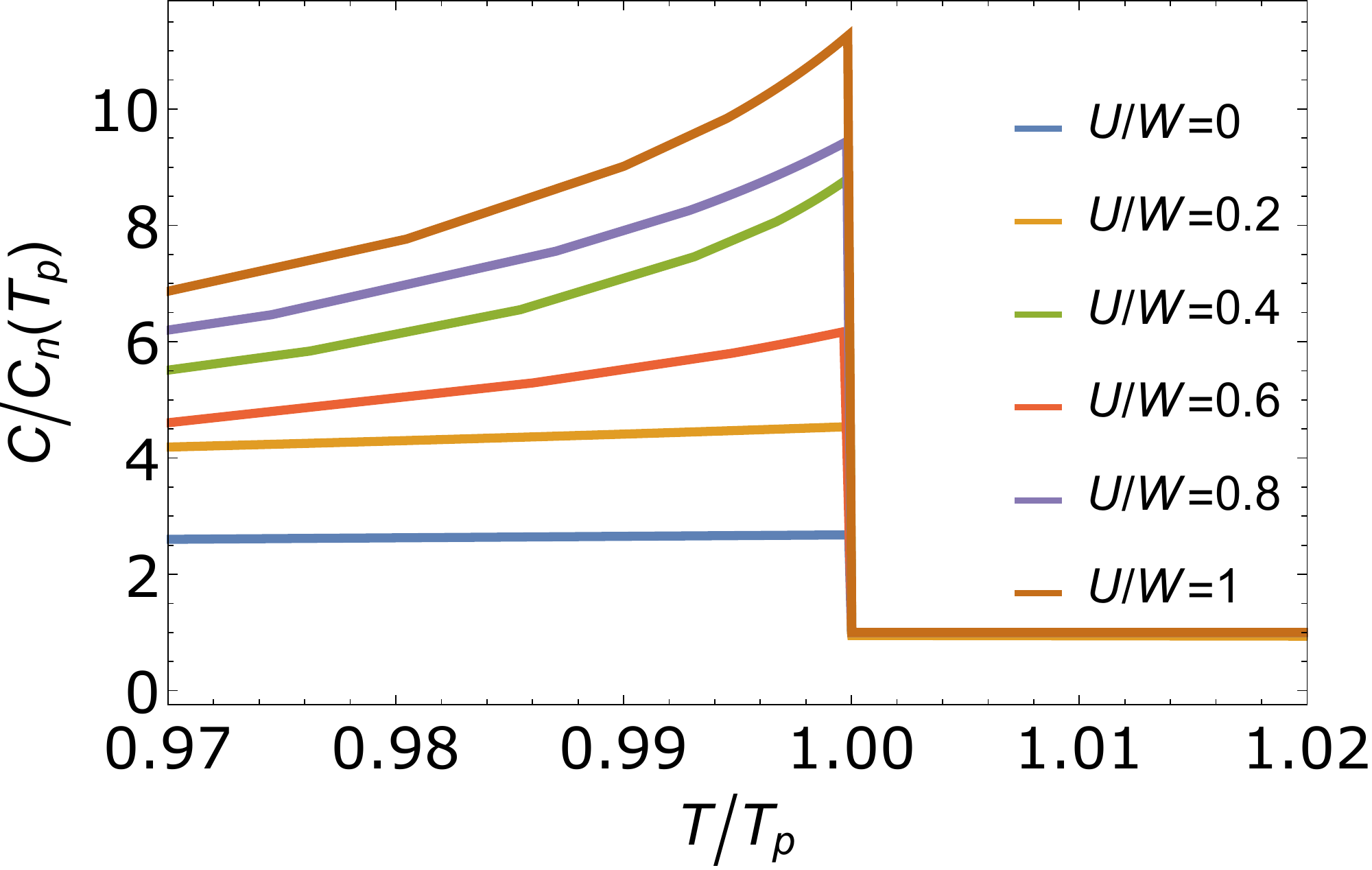}
 \caption{
 Heat capacity $C$ near the phase transition at $T = T_p$, at pair coupling $g/W = 0.53$ and for representative values of the Mott coupling $U$.
 $C$ is normalized by its value $C_n(T_p)$ on the normal side of the phase transition.
 }
 \label{fig:heat-capacity}
\end{figure}

In the HKSC where the gap changes discontinuously, the heat capacity still undergoes such a jump discontinuity, although the size of the discontinuity $\Delta C$ no longer scales universally with $C_n(T_c)$.  Rather as we will see, it depends on the Mott scale $U$ as well.
When the gap changes discontinuously with temperature, the entropy also changes discontinuously so that the heat capacity $C = T \partial_T S$ is singular at the transition.
We show its behavior in Fig.~\ref{fig:heat-capacity}, omitting the singularity.  As is evident,  Mottness, as tracked by increasing $U$, enhances the heat capacity jump at $T_p$, the temperature of the first-order transition.  Similar enhancements will be seen as well for the ultrasonic attenuation and the spin-lattice relaxation rate.

\subsection{NMR}

The key feature of BCS is computability with the mean-field formalism. The same can be done here as we know all the excited states and hence can calculate all of the experimental quantities delineated by BCS. As an example, we compute the spin-lattice relaxation rate which in a BCS superconductor exhibits a peak\cite{HSpeak} below $T_c$. 
The spins of atomic nuclei relax by exchanging energy with their environment. As the spins in a superconductor are in phase, there is an enhancement below $T_c$. The relaxation rate $1/T_1$ of nuclei in an electronic environment is related to the transverse dynamic spin susceptibility of the quasiparticles, by\cite{coleman2015introduction}
\begin{equation}
 \frac{1}{TT_1}=\lim_{\omega\rightarrow0}\frac{2k_B}{\gamma_e^2\hbar^4}\sum_q|A_H(q)|^2\frac{\Im{\chi^{s}}(q,\omega)}{\omega}
\end{equation}
where $\gamma_e$ is the electron gyromagnetic ratio, $A_H(q)$ is the hyperfine coupling of the contact interaction with electron spins, and $\Im{\chi^s}$ is the imaginary part of the spin susceptibility.\par
In a normal metal state, $\frac{\Im{\chi^{s}}(\omega)}{\omega}\sim N(0)^2$. This leads to a linear dependence of the nuclear relaxation rate on temperature, referred to as \emph{Korringa relaxation} law\cite{korringa1950nuclear}
\begin{equation}
 \frac{1}{TT_1}\sim \frac{2k_B}{\gamma_e^2\hbar^4}N(0)^2\sum_q|A_H(q)|^2={\rm constant}.
\end{equation}
\begin{figure}[h]
 \centering
 \includegraphics[width=8.6cm]{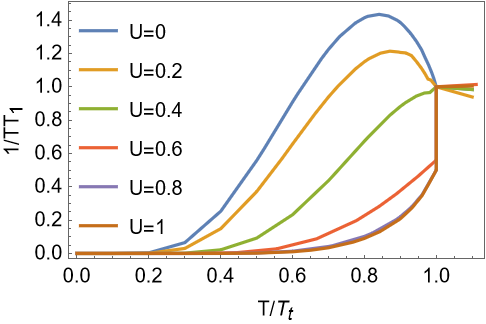}
 \caption{The NMR relaxation rate $1/TT_1$(normalized by the $1/TT_1$ value at the transition temperature ($T=T_t$) in normal state) as a function of temperature at superconducting pairing strength $g=0.53$, with varying value of $U$. The temperature is scaled by the gap-oppening temperature, $T_t=T_2$(second-order transition) for $U=0,U=0.2$, and $T_t=T_p$(first order transition) for others. To regulate the divergence of the delta function when doing momentum summation, we introduced a small imaginary damping rate $i\epsilon$ to the frequency where $\epsilon/W=0.01$.}
 \label{fig:NMR}
\end{figure}
In a BCS superconductor, we need to take account of the strongly energy-dependent quasiparticle density of states $N(E)\rightarrow N(0)\frac{|E|}{\sqrt{E^2-\Delta^2}}$. If we further assume a point contact interaction , $A(q)=A$ is a constant with respect to $q$, the relaxation rate becomes
\begin{equation}
 \frac{1}{TT_1}\propto \int_\Delta^\infty dE(\frac{df}{dE})\frac{E^2}{E^2-\Delta^2},
\end{equation}
which generates the Hebel-Slichter peak right below the transition temperature $T_c$, which is shown in Fig. \ref{fig:NMR} as $U=0$(blue curve).\par

By contrast, in the HKSC model, there is no fermionic quasipartical excitation of the ground state. As a result, we shall calculate the susceptibility $\chi^s(q,\omega)$ from its basic definition. The dynamical susceptibility in imaginary time is
\begin{equation}
\begin{split}
 \chi^s_{ab}(q,i\nu_n)&=\ave{M_a(q)M_b(-q)}\\
 &=\int_0^\beta\ave{M_a(q,\tau)M_b(-q,0)}e^{i\nu_n\tau}.
\end{split}
\end{equation}
For a spin-isotropic system, we have $\chi^s_{ab}(q)=\delta_{ab}\chi^s(q)$.
Thus, we can calculate the $z$-axis response to a field appllied along $z$: $\chi^s(q)=\ave{M_z(q)M_z(-q)}$. Since
\begin{equation}
 M_z(q)=\sum_{\vec k}(c_{\vec {k-q},\uparrow}^\dagger c_{\vec k,\uparrow}-c_{\vec {k-q},\downarrow}^\dagger c_{\vec k,\downarrow}),
\end{equation}
we find that
\begin{equation}
\begin{split}
 \chi^s(q,\tau)&=\ave{M_z(q,\tau)M_z(-q,0)}=\chi_r(q,\tau)-\chi_a(q,\tau),
\end{split}
\end{equation}
where 
\begin{equation}
\begin{split}
\chi_{r}(q, \tau) &=\sum_{\vec k, \sigma}\ave{ c_{\vec {k-q}, \sigma}^{\dagger}(\tau) c_{\vec k, \sigma}(\tau) c_{\vec k, \sigma}^{\dagger}(0)c_{\vec {k-q}, \sigma}(0) }\\
\chi_{a}(q, \tau) &=\sum_{\vec k, \sigma}\ave{ c_{\vec {k-q}, \sigma}^{\dagger}(\tau) c_{\vec k, \sigma}(\tau) c_{\vec{-k+q}, \bar\sigma}^\dagger(0) c_{\vec{-k}, \bar\sigma}(0)}
\end{split}
\end{equation}
are the regular and anomalous parts of the correlation function. The anomalous term is non-zero only if the ground state no longer conserves the particle number as in the superconducting phase.
The imaginary part of spin susceptibility takes the form
\begin{widetext}
\begin{equation}
\begin{split}
 \mathrm{Im}\chi(q,\nu-i0^+)=\sum_{\vec k, \sigma} \sum_{i, j, i^{\prime}, j^{\prime}}\left(p_{\vec k}^{j} p_{\vec {k-q}}^{i^{\prime}}-p_{\vec k}^{i} p_{\vec {k-q}}^{j^{\prime}}\right)
 \left[\left|M_{\vec k, \sigma}^{i j}\right|^{2}\left|M_{\vec {k-q}, \sigma}^{i^{\prime} j^{\prime}}\right|^{2}+M_{\vec k, \sigma}^{i j} M_{-\vec k, \bar{\sigma}}^{j i} \bar{M}_{\vec {k-q}, \sigma}^{i^{\prime} j^{\prime}} \bar{M}_{-\vec{k+q}, \bar{\sigma}}^{j^{\prime} i^{\prime}}\right]
 \delta(\nu-(\omega_{\vec k}^{ji}-\omega_{\vec {k-q}}^{j^\prime i^\prime})),
\end{split}
\label{eq:NMR}
\end{equation}
\end{widetext}
where $(E_{\vec k}^i,\ket{\psi_{\vec k}^i})_{i=1,...,16}$ is the eigensystem of $H_{\vec k}$, $p_{\vec k}^i=e^{-\beta E_{\vec k}^i}/Z_{\vec k}$ is a Boltzmann weight, $\omega_{\vec k}^{ij}=E_{\vec k}^i-E_{\vec k}^j$ is an excitation energy, and $M_{\vec k\sigma}^{ij}=\ave{\psi_{\vec k}^i|c_{\vec k\sigma}|\psi_{\vec k}^j}$. Thus we can write the NMR relaxation rate as
\begin{equation}
\begin{split}
 \frac{1}{TT_1}&\propto \sum_q \left. \frac{\Im[\chi(q,\nu-i\delta)]}{\nu}\right|_{\nu\rightarrow0}\\
 &=\sum_q\sum_{\vec k,\sigma}\sum_{i, j, i^{\prime}, j^{\prime}}(p_{\vec k}^i+p_{\vec k}^j)(p_{\vec {k-q}}^{i^{\prime}}+p_{\vec {k-q}}^{j^{\prime}})\\ &\qquad f'(\omega_{\vec k}^{ij})M_{\vec k,\vec q,\sigma}^{i, j, i^{\prime}, j^{\prime}}\delta(\omega_{\vec k}^{ji}-\omega_{\vec {k-q}}^{j^\prime i^\prime}),
\end{split}
\label{eq:small-nu}
\end{equation}
where
$M_{\vec k,\vec q,\sigma}^{i, j, i^{\prime}, j^{\prime}}$ is the term inside the square braket from Eq.(\ref{eq:NMR}) and $f(x)=1/({e^{\beta x}+1})$ is the Fermi-distribution function.
The NMR relaxation rate of the HKSC model can be calculated numerically. Thus we use the expression given above together with the self-consistent pair amplitude to obtain the temperature dependence of $1/TT_1$.
Fig. \ref{fig:NMR} showes the temperature dependence of $1/TT_1$ at $g=0.53$. For $U<0.4$, we have a second-order superconducting transtion and as $U$ increases, the HS peak shrinks and is completely absent when The first order transtion takes place at $U\ge 0.4$. At sufficiently low temperature, $1/TT_1$ has an exponential dependence on temperature. Once again, we see that Mottness is the culprit in leading to significant deviations from the standard BCS theory. Previous work\cite{HSpeak2} has attributed the absence\cite{NMR1,NMR2,NMR3} of the HS peak in the cuprates to spin fluctuations. In the current work, such fluctuations are absent. Hence, we advocate that Mott physics alone, the tendency to have single occupancy below the chemical potential, is sufficient to kill the Hebel-Slichter peak.

\section{Ultrasonic attenuation}

Alongside the nuclear spin relaxation rate, another standard observable that probes the superconducting gap is the attenuation rate $\alpha$ of ultrasonic phonons transmitted through the sample.
The attenuation rate,
\begin{equation}
 \alpha(q) = \Gamma(q, \nu = \omega_q),
\end{equation}
of phonons with momentum $q$ is given by the on-shell decay rate $\Gamma(q, \nu = \omega_q)$ of the phonon propagator, where $\omega_q \sim v_\ph q$ is the phonon dispersion at small momenta $q$.
Typical experiments are conducted with phonon frequencies on the order of $f \sim 10^7 \mathrm{Hz}$~\cite{morse1957}.
Supposing a speed of sound $v_\ph \sim 10^4 \mathrm{m/s}$ in the sample, these phonons have energy $E_\ph/k_B = \hbar 2\pi f /k_B \sim 5 \x 10^{-4} \mathrm{K}$ and momentum $q = 2\pi f/s \sim 6 \x 10^{-7} \text{\AA}\inv$.
Since $E_\ph/k_B \ll T_c$ and $q \ll k_F \sim \text{\AA}\inv$, we focus on the limit
\begin{equation}
 \alpha \equiv \alpha(q \to 0).
\end{equation}
At weak electron-phonon coupling, the decay rate
\begin{equation}
 \Gamma(q,\nu) = \Im\chi^{(c)}(q,\nu-i0^+)
\end{equation}
is given in turn by the electronic charge susceptibility
\begin{equation}
 \chi^{(c)}(q,i\nu_n)
 \equiv \int_0^\beta d\tau\; e^{i\nu_n \tau} \chi^{(c)}(q, \tau)
\end{equation}
evaluated at the bosonic Matsubara frequency $\nu_n = 2\pi n/\beta$.
In imaginary time,
\begin{align}
 &\quad \chi^{(c)}(q, \tau) \NN\\
 &\equiv \sum_{\vec k,k',\sigma,\sigma'} \ave{c_{\vec {k-q},\sigma}\adj(\tau) c_{\vec k,\sigma}(\tau) c_{\vec k'+q,\sigma'}\adj(0) c_{\vec k',\sigma'}(0)}_{0,c} \\
 &= \chi_r(q, \tau) + \chi_a(q, \tau),
\end{align}
where $\chi_r$ and $\chi_a$ are the regular and anomalous correlation functions introduced in the previous section.
The sign between the two terms is reversed relative to the spin susceptibility. 
This prevents any coherence peak in the ultrasonic attenuation rate here, just as it does for BCS superconductors.
At small momenta $q$, the attenuation rate takes the form
\begin{align}
 \alpha(q)
 &\sim \omega_q \sum_{\vec k,\sigma} \sum_{i,j,i',j'}
 (p_{\vec k}^i+p_{\vec k}^j) (p_{\vec k}^{i'}+p_{\vec k}^{j'})\\
 &\quad \times f'(\omega_{\vec k}^{ji})
 M_{\vec k,0,\sigma}^{i,j,i',j'} \delta\P{v_\ph q - (\omega_{\vec k}^{ji}-\omega_{\vec {k-q}}^{j'i'})}, \NN
\end{align}
analogous to Eq.~(\ref{eq:small-nu}) for the spin relaxation rate.
Its convergence in the limit $q \to 0$ can be seen as follows for the standard context~\cite{tsuneto1961}, i.e.\ in $d=3$ dimensions and with an isotropic quadratic dispersion $\xi_k = k^2/2m-\mu$.
The sum is dominated by the excitation between the ground state and the lowest-lying odd-parity level, for which $(i,j) = (i',j')$ and the delta function resolves to $\delta(v_\ph q - \vec{q} \cdot \grad_\vec{k} \omega_\vec{k}^{ji})$.
Pulling this back to a form that can be formally integrated over $k$, i.e. $\delta(k - k_0)/\abs{\cdots}$, then extracts a factor of $1/q$, resulting in an overall $q$-dependence given by the product $\omega_q/q \sim v_\ph$.

\begin{figure}[htpb]
 \centering
 \includegraphics[width=\linewidth]{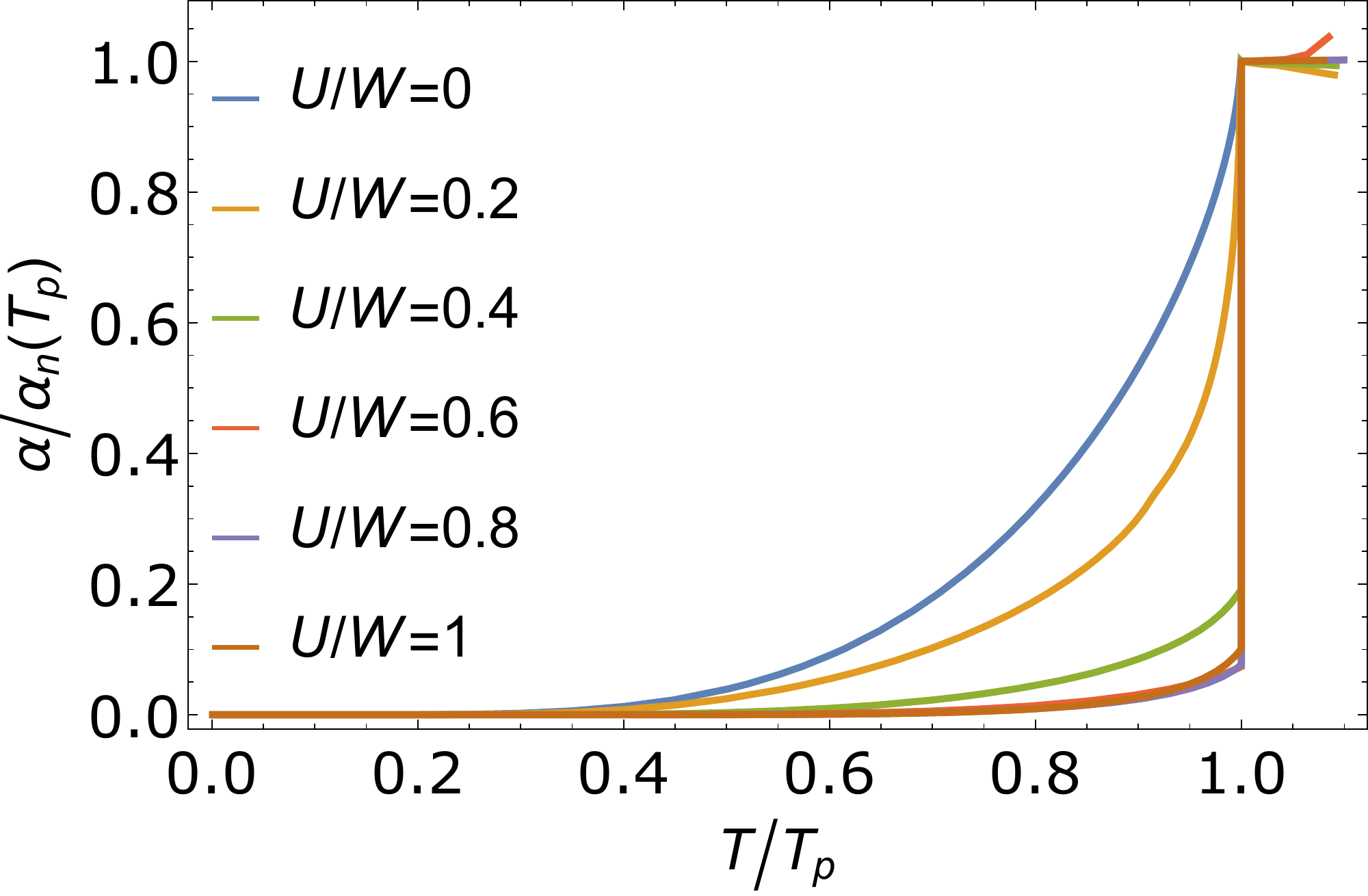}
 \caption{
 Ultrasonic attenuation rate $\alpha$ in the superconducting phase, at pair coupling $g/W = 0.53$ and for representative values of the Mott coupling $U$.
 $\alpha$ is normalized by its value $\alpha_n(T_p)$ on the normal side of the phase transition at $T = T_p$.
 The curves for $U/W = 0.6$ and $0.8$ are difficult to see here because they are covered by the $U/W=1$ curve.
 }
 \label{fig:USA}
\end{figure}

For general parameters the attenuation rate $\alpha$ must be evaluated numerically, as shown in Fig.~\ref{fig:USA}.
In line with the NMR relaxation rate calculation, we take an isotropic dispersion that is cut off at some magnitude of the crystal momentum.
Unlike the NMR calculation, however, the probe momentum $q$ is taken asymptotically to zero instead of being summed over, so it is necessary in this case to perform the integral directly in momentum space.
For the figure,  we have used $v_\ph/(\mathsf{a}W) = 10^{-2}$ and $q = 10^{-7}\pi/\mathsf{a}$.
The resulting attenuation rate $\alpha$ decreases monotonically in the superconductor from the normal-phase value $\alpha_n(T_p)$, changing discontinuously at the phase transition when the gap $\Delta$ opens discontinuously.
At low temperatures it decays exponentially as $e^{-\Delta/T}$.
This allows the gap to be extracted from the ultrasonic attenuation rate in the HKSC.  Once again, we see that as the overall precipitous fall-off of the ultrasonic attenuation rate is accented as the  Mott parameter $U$ increases.  In principle, this trend is experimentally testable.

\section{Final Remarks}
\begin{table*}[ht]
\caption{Summary of the superconducting properties in the  HK model and a Fermi liquid(FL). $\chi$ represents the pair susceptibility, $\Delta$ is the pairing gap,  H. S. stands for the Hebel-Slichter\cite{HS} peak, a key feature of BCS superconductors.}
 \centering
 \begin{tabular}{|c|c|c|}
  \hline
  property& FL& Mottness (HK) \\
  \hline
  $\chi$ divergent at& $T_c$ & $T_c(=T_2)$\\
  $\Delta$ opens at & $T_c$ & $T_p(>T_2)$\\
  $\lim_{g\rightarrow0} 2\Delta_0/k_BT_c$ & 3.53 & $\infty$\\
  $E_{\rm cond}/N(0)\Delta^2$ & -1 & $[-2,-1]$\\
  Quasiparticles & Bogoliubons & PHYons\footnote{linear combinations of Holon and Doublon\cite{hksupercon}}\\
  Ginzburg reduced $t_G$& $\sim10^{-9}$&$\sim10^{-11}$\\
  NMR $1/TT_1$ & H-S peak & no H-S peak\\
  Landau Expansion& $a=\alpha t$ \footnote{$t=\frac{T-T_c}{T_c}$ and $\alpha>0$ for both cases}$,b>0$ &$a=\alpha t,b<0, c>0$\\
  Ultrasonic attenuation, $\alpha$ & $\sim e^{-\Delta/T}$ below $T_c$ & $\sim e^{-\Delta/T}$ below $T_p<T_c$ \\
  \hline
 \end{tabular}

 \label{tab:my_label}
\end{table*}
The primary difficulty in unlocking how superconductivity arises in a doped Mott insulator is computation with a controlled theory. The HK model enables such an analysis as it represents a fixed point for quartic Fermionic theories that break the $\mathbb Z_2$ symmetry of Fermi liquids. Since this includes the Hubbard model, it suffices to analyse the tractable HK model. An analogy with Fermi liquids is relevant here. The relevant physics of a Fermi liquid follows from the free quadratic theory of Eq. (\ref{free}) as all short-range repulsions are irrelevant. That such repulsions are irrelevant follows from the simple fact that Fermi liquids are local in momentum space. Destruction of this state, except for pairing, requires an equally local interaction in momentum space. The HK interaction is just the most relevant interaction in momentum space that suffices as it maximally breaks the $\mathbb Z_2$ symmetry of a Fermi liquid.
Carrying out a pairing analysis from this starting point should reveal the key differences with how superconductivity obtains from doped Mott insulator as opposed to a Fermi liquid.

Table (\ref{tab:my_label}) catalogs the differences between superconductivity from a FL with the Mott counterpart.  As determined here, the first key dfference is the appearance of two energy scales, the pairing temperature, $T_p$ and the temperature at which the susceptiblity diverges.  In the absence of Mottness, only a single scale characterizes superconductivity.  Nonetheless, in the HK model we still find that the mean-field theory is essentially exact as the Ginzburg reduced temperature is vanishingly small. A key prediction here is that Mottness makes the underlying transition first order.  This can be confirmed by careful mesurements of the latent heat in the cuprates.  Another key prediction here is that the Hebel-Slichter peak is killed by the strong correlations of the Mott state. While it had been speculated that antiferromagnetic correlations diminish the relaxation rate\cite{HSpeak2}, what we find here is that in a model that has Mott physics but no antiferromagnetism, the HS peak does not survive. Experimentally, the best NMR data\cite{NMR1,NMR2,NMR3} indicate that on the Cu or O sites, no Hebel-Slichter peak exists.  Our work indicates that the suppression of the HS peak is due entirely the bifurcation of the spectrum into upper and lower Hubbard bands.  Such a bifurcation prevents the coherence that is typically thought to be the mechanism behind the HS peak.  It is these strong correlations of the Mott state that lead to a deviation as well from the standard Bogoliubov quasiparticles and the onset of the composite excitations, PYHons\cite{hksupercon}, as the new quasi-excitations above the superconducting ground state.  Such correlations also enhance the condensation energy and lead to a divergence of $\lim_{T\rightarrow 0} 2\Delta_0/T_c$ in the HK superconductor.  As all of these trends are traceable to the strong correlations of the Mott state, we conclude that Table (\ref{tab:my_label}) should provide the blueprint for superdconductivity in doped Mott insulators.

\textbf{Acknowledgements} 

PWP and JZ thank DMR-2111379 for partial funding of this project. E.W.H. was supported by the Gordon and Betty Moore Foundation EPiQS Initiative through the grants GBMF 4305 and GBMF 8691.
\section*{Appendix}
\subsection{stiffness}
The superfluid stiffness is defined by
\begin{equation}
    F[\theta]=F[0]+\int d^dx\frac{1}{2}\rho_s(\nabla\theta)^2,
\end{equation}
where $F=-\frac{1}{\beta}\log Z$ is the free energy and $\theta$ is the phase of the superfluid.
Consider applying the following twist
\begin{equation}
    c_i^\dagger\rightarrow c_i^\dagger e^{i\phi r_{i,x}}.
    \label{eq:twist}
\end{equation}
For charge-$n$e superconductivity, $\nabla\theta=n\phi\hat{x}$ and we can calculate the stiffness by
\begin{equation}
    \begin{split}
        \rho_s&=\frac{1}{n^2}\frac{1}{N}\left.\frac{\partial^2F[\phi]}{\partial\phi^2}\right|_{\phi=0}\\
        &=-\frac{1}{n^2}\frac{1}{N}\frac{1}{\beta}\left[\frac{\partial_\phi^2Z}{Z}-\left(\frac{\partial_\phi Z}{Z}\right)^2\right]_{\phi=0}.
    \end{split}
\end{equation}
Here we have set the lattice constant to 1, and $N$ is the total number of unit cells.
In order to derive the correct stiffness for an HK superconductor, we work with the Hamiltonian
\begin{equation}
    H=\sum_{\vec k\sigma}\epsilon_{\vec k}c_{\vec k\sigma}^\dagger c_{\vec k\sigma}+\sum_{\vec k}Un_{\vec k\uparrow}n_{\vec k\downarrow}+H_p,
\end{equation}
where $H_p$ is the general superconducting pairing term
\begin{equation}
    H_p=g\sum_{\vec q}\Delta_{\vec q}^\dagger\Delta_{\vec q}
\end{equation}
and $\Delta_{\vec q}=\sum_{\vec k}c_{\vec k+\vec q\uparrow}c_{-\vec k\downarrow}$ is the copper pair creation operator with momentum $q$.
Under the twist we introduced in Eq.(\ref{eq:twist}), the fourier transform of the fermion operator becomes
\begin{equation}
\begin{split}
    c_{\vec k}^\dagger&=\frac{1}{\sqrt{N}}\sum_ic_i^\dagger e^{i\vec k\cdot \vec r_i}\\
    &\rightarrow \frac{1}{\sqrt{N}}\sum_ic_i^\dagger e^{i\vec k\cdot \vec r_i+i\phi r_{i,x}}\\
    &=c_{\vec k+\phi \hat{e}_x}^\dagger.
\end{split}
\end{equation}
Thus the Hamiltonian under the twist
\begin{align}
    \begin{split}
        H[\phi]&=\sum_{\vec k\sigma}\epsilon_{\vec k}c_{\vec k+\phi \hat{e}_x\sigma}^\dagger c_{\vec k+\phi \hat{e}_x\sigma}\\
        &+\sum_{\vec k}Un_{\vec k\uparrow}n_{\vec k\downarrow}+H_p[\phi]
    \end{split}\\
    H_p[\phi]&=g\sum_{\vec q}\Delta_{\vec q}^\dagger[\phi]\Delta_{\vec q}[\phi]\\
    \Delta_{\vec q}[\phi]&=\sum_{\vec k} c_{\vec k+\vec q+\phi \hat{e}_x\uparrow}c_{-\vec k+\phi \hat{e}_x\downarrow}=\Delta_{\vec q+2\phi \hat{e}_x},
\end{align}
where the HK term and the superconducting pairing terms do not change under the twist. Thus,
\begin{equation}
    H[\phi]-H[0]=\sum_{\vec k\sigma}(\epsilon_{\vec k}-\epsilon_{\vec k-\phi\hat{e}_x})c_{\vec k\sigma}^\dagger c_{\vec k\sigma}.
\end{equation}
We expand the Hamiltonian to second-order in $\phi$ to obtain
\begin{eqnarray}
    H[\phi]-H[0]&=-\phi J_x+\frac{1}{2}\phi^2T_x\\
    J_x&=\sum_{\vec k\sigma}(\partial_{ k_x}\epsilon_{\vec k})c_{\vec k\sigma}^\dagger c_{\vec k\sigma}\\
    T_x&=\sum_{\vec k\sigma}(\partial^2_{ k_x}\epsilon_{\vec k})c_{\vec k\sigma}^\dagger c_{\vec k\sigma},
\end{eqnarray}
with $J_x$  the total current in the $x$ direction and $T_x$ is the kinetic energy arising from hopping in the $x$ direction. The partition function and its derivatives are
\begin{align}
    \begin{split}
        Z[\phi]&=\tr{e^{-\beta H[\phi]}}\\
        &=\tr{e^{-\beta H[0]}Te^{\int_0^\beta d\tau\phi J_x(\tau)-\frac{1}{2}\phi^2T_x(\tau)]}}
    \end{split}\\
    \begin{split}
        \left.\frac{\partial_\phi Z[\phi]}{Z}\right|_{\phi=0}&=\frac{1}{Z}\tr{e^{-\beta H[0]}T\left[\int_0^\beta d\tau J_x\right]}\\
        &=\beta\ave{J_x}=0
    \end{split}\\
    \begin{split}
        \left.\frac{\partial^2_\phi Z[\phi]}{Z}\right|_{\phi=0}&=\frac{1}{Z}\tr{e^{-\beta H[0]}T\left[\left(\int_0^\beta d\tau J_x\right)^2-\int_0^\beta d\tau T_x\right]}\\
        &=\beta\left(\int_0^\beta d\tau\ave{J_x(\tau)J_x}\right)-\beta\ave{T_x}.
    \end{split}
\end{align}
Finally, the superfluid stiffness is
\begin{equation}
    \rho_s=\frac{1}{n^2N}\left(\ave{T_x}-\int_0^\beta d\tau\ave{J_x(\tau)J_x}\right).
\end{equation}
The decrease of the superfluid stiffness in the HK model\cite{hksupercon} was calculated at $T=0$. For finite temperature, especially near $T_c$, we expect that the ratio $\rho_s/\Delta^2$ remains finite.
\begin{figure}[h]
 \centering
 \includegraphics[width=8.6cm]{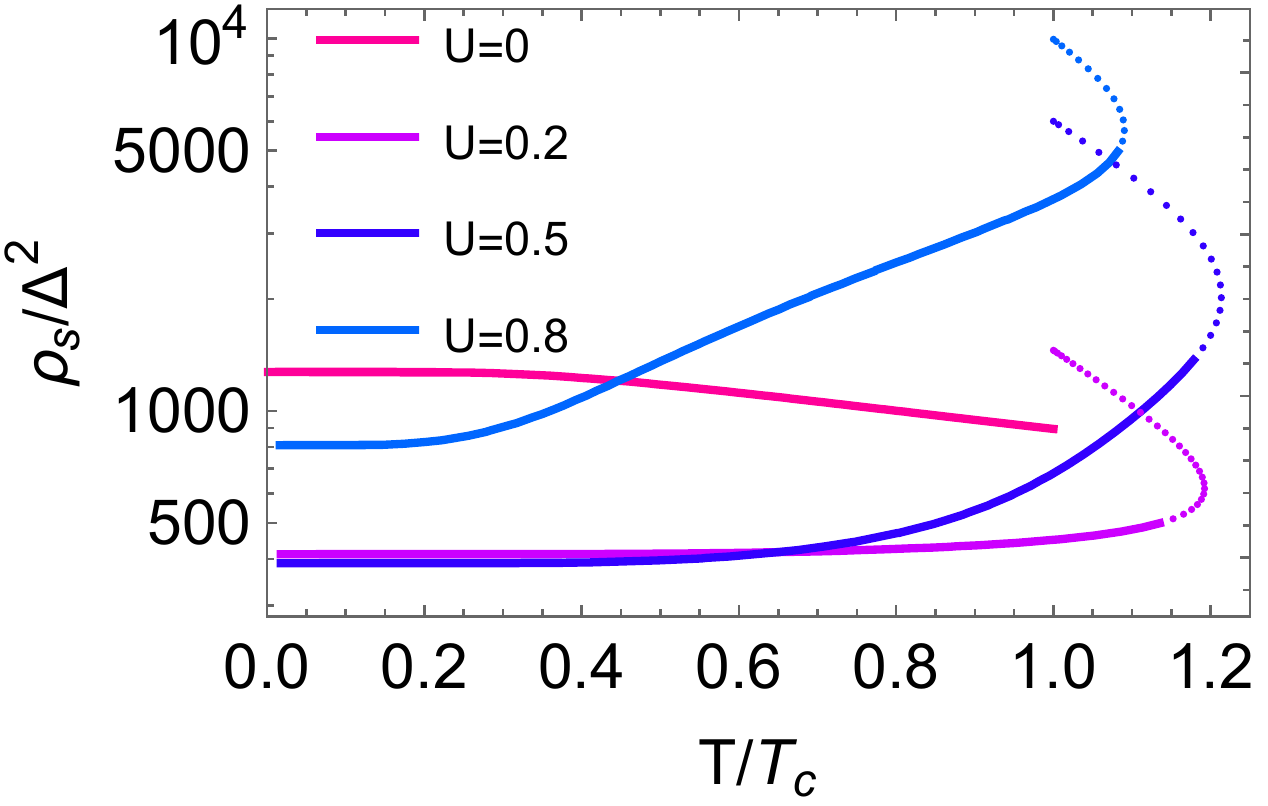}
 \caption{Finite temperature superfluid stiffness. For the sake of illustrating the ginzburg criterion, we keep the multiple solutions to the self-consistant gap equation even though only one of them represents the true global minimum of the free energy. The first order transition happens at $T_p$ where the solid lines change into dotted points for $U=0.2,0.5,0.8$.}
 \label{fig:fintstiffness}
\end{figure}
Fig. \ref{fig:fintstiffness} shows the temperature dependence of the superfluid stiffness for multiple value of $U/W$. Near zero temperature the HK pairing reduce the stiffness by approximately 2 times. The stiffness close to $T_c$, however, was increased drastically up to 10 times. The increased stiffness guarantees that the Ginzburg reduced temperature is small in the HK model, and thus proved the applicability of the mean-field theory of HK superconducting model.
\clearpage

\bibliography{hkthermo}

\end{document}